\newif\ifHEVEA
\newif\ifDRAFT
\newif\ifFINAL
\newif\ifREVWS
\newif\ifEXPLE
\newif\ifACCEPTED

\HEVEAfalse
\DRAFTfalse
\REVWSfalse
\EXPLEfalse
\FINALtrue
\ACCEPTEDtrue

\documentclass[acmlarge,screen,nonacm]{acmart}

\newcommand{\abstractcount}[1]{}

\def\tablefontsize{\small}

\newif\ifSTATSA \newif\ifSTATSB \newif\ifPERA \newif\ifPERB \newif\ifPERC \newif\ifHOURA \newif\ifHOURB \newif\ifHOURC 

\STATSAfalse
\STATSBfalse

\PERAfalse
\PERBfalse
\PERCfalse

\HOURAfalse
\HOURBfalse
\HOURCfalse

\def\CS{CairnSCREEN}
\def\CF{CairnFORM}
\def\PS{PlantSCREEN}
\def\PF{PlantFORM}

\def\raL{\raggedleft}
\def\raR{\raggedright}

\def\dg{$\dagger$}

\newcommand\ts[1]{\textsubscript{#1}}

\newcommand\tbh[1]{\textbf{\textit{#1}}}

\newcommand\mcli[1]{\multicolumn{1}{l}{\textit{#1}}}

\usepackage{xcolor}

\newcommand{\magenta}[1]{#1}

\definecolor{darkgreen}{rgb}{0.0, 0.4, 0.13}

\usepackage[normalem]{ulem}

\newcommand\xoutpars[1]{\let\helpcmd\xout\parhelp#1\par\relax\relax}
\newcommand\soutpars[1]{\let\helpcmd\sout\parhelp#1\par\relax\relax}
\long\def\parhelp#1\par#2\relax{\helpcmd{#1}\ifx\relax#2\else\par\parhelp#2\relax\fi }

\usepackage{soul}

\setcitestyle{nocompress}

\usepackage{pdfpages}

\usepackage{graphicx}
\usepackage{float}
\usepackage{threeparttable}
\usepackage{tabularx}
\usepackage{multirow}
\usepackage{caption} 
\usepackage{subcaption}
\usepackage[export]{adjustbox}

\usepackage{cleveref}

\crefname{figure}{Figure}{Figures}

\newcommand{\killpunct}[1]{}

\definecolor{ACMBlue}{cmyk}{1,0.1,0,0.1}
\definecolor{ACMYellow}{cmyk}{0,0.16,1,0}
\definecolor{ACMOrange}{cmyk}{0,0.42,1,0.01}
\definecolor{ACMRed}{cmyk}{0,0.90,0.86,0}
\definecolor{ACMLightBlue}{cmyk}{0.49,0.01,0,0}
\definecolor{ACMGreen}{cmyk}{0.20,0,1,0.19}
\definecolor{ACMPurple}{cmyk}{0.55,1,0,0.15}
\definecolor{ACMDarkBlue}{cmyk}{1,0.58,0,0.21}

\definecolor{ACMBLUE}{cmyk}{1,0.1,0,0.1}
\definecolor{ACMYELLOW}{cmyk}{0,0.16,1,0}
\definecolor{ACMORANGE}{cmyk}{0,0.42,1,0.01}
\definecolor{ACMRED}{cmyk}{0,0.90,0.86,0}
\definecolor{ACMLIGHTBLUE}{cmyk}{0.49,0.01,0,0}
\definecolor{ACMGREEN}{cmyk}{0.20,0,1,0.19}
\definecolor{ACMPURPLE}{cmyk}{0.55,1,0,0.15}
\definecolor{ACMDARKBLUE}{cmyk}{1,0.58,0,0.21}
\definecolor{BLACK}{rgb}{0,0,0}

\newcommand\reduline{\bgroup\markoverwith
{\textcolor{gray}{\rule[0.5ex]{2pt}{0.8pt}}}\ULon}
 
\newcommand{\erase}[1]{}

\AtBeginDocument{}

\setcopyright{cc}
\copyrightyear{2026}
\acmYear{2026}
\acmDOI{XXXXXXX.XXXXXXX}

\begin{document}

\title[Shaping Plant-Like Shape-Changing Interfaces as Vertical Charts]{Shaping Plant-Like Shape-Changing Interfaces as Vertical Charts: Maximizing Readability, Aesthetics, and Naturalness}

\author{\'Elodie Bouzekri}
\authornote{Both authors contributed equally to this research and to this paper.}
\email{elodie.bouzekri@univ-brest.fr}
\affiliation{\institution{Univ. Bordeaux, ESTIA-Institute of Technology, EstiaR}
\city{Bidart}
  \country{France}
  \postcode{F-64210}
}
\affiliation{\institution{Lab-STICC UMR 6285, Univ. Brest}
\city{Brest}
  \country{France}
  \postcode{F-29238}
}
\orcid{0000-0002-7902-3541}
\author{Guillaume Rivi\`ere}
\orcid{0000-0001-8390-9751}
\authornotemark[1]
\email{g.riviere@estia.fr}
\affiliation{\institution{Univ. Bordeaux, ESTIA-Institute of Technology, EstiaR}
\city{Bidart}
  \country{France}
  \postcode{F-64210}
}

\renewcommand{\shortauthors}{Bouzekri and Riviere}

\begin{abstract}
Conveying environmental data has grown interest in encouraging the adoption of eco-friendly lifestyles through data-driven strategies. This scope appeals to data visualizations representing the environmental purpose. For example, previous work has already proposed nature-inspired counters, gauges, and bitmaps, but data series remains to be explored. Therefore, could we design and implement effective plant-like charts? This paper brings answers through a research-through-design approach that explores a design space to maximize readability and aesthetics. It then compares four prototypes of charts over modality and material dimensions by asking users about scenarios involving renewable energy forecasts. The results examine whether implementing physical charts is worth it instead of graphical charts and the advantages of using meaningful materials that evocate sustainability and enhance naturalness.
The results also reexamine, with physical charts, the previous results on graphical infographics of slightly lower clarity and readability but higher aesthetics of embellishment.  
In addition, learnability is examined for encoding rates through folded shapes. This paper shows that physical plant-like charts are worthwhile because of promising performance and best-of-breed naturalness when materials allow low-tech aspects' perception and because being installable in public places without explanations if folded shapes encode rates ranging from 0 to a maximum value.
\end{abstract}




\def\protoH{4.2cm} \def\figWratio{0.146}
\def\spaceW{3mm}

\begin{teaserfigure}
   \centering
  \begin{subfigure}[b]{\figWratio\linewidth}
    \centering
    \includegraphics[height=\protoH]{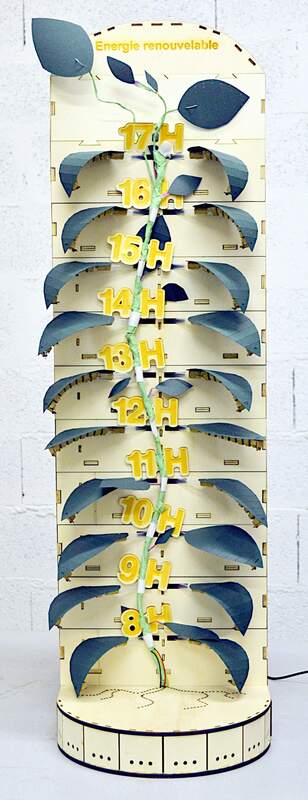} \caption{From 8:00 to 17:59, peak at 12H.}
    \label{fig:teaser:var1}
  \end{subfigure}\hspace{\spaceW}
  \begin{subfigure}[b]{\figWratio\linewidth}
    \centering
    \includegraphics[height=\protoH]{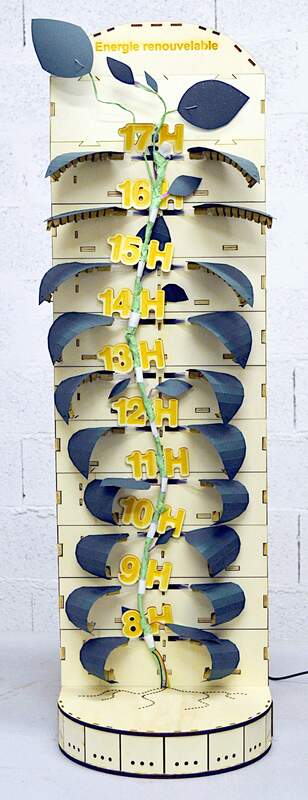} \caption{From 10:00 to 17:59, peak at 16H.}
    \label{fig:teaser:var2}
  \end{subfigure}\hspace{\spaceW}
  \begin{subfigure}[b]{\figWratio\linewidth}
    \centering
    \includegraphics[height=\protoH]{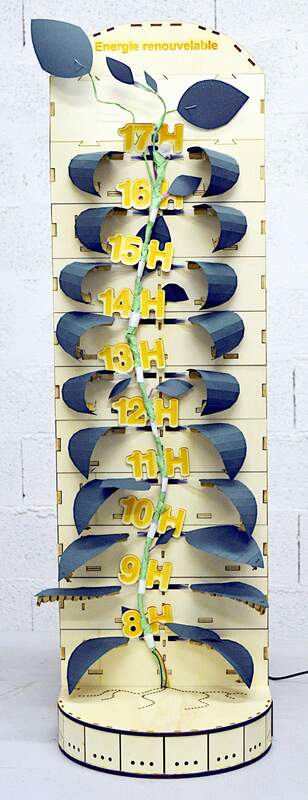} \caption{From 8:00 to 12:59, peak at 9H.}
    \label{fig:teaser:var3}
  \end{subfigure}\hspace{\spaceW}
  \begin{subfigure}[b]{\figWratio\linewidth}
    \centering
    \includegraphics[height=\protoH]{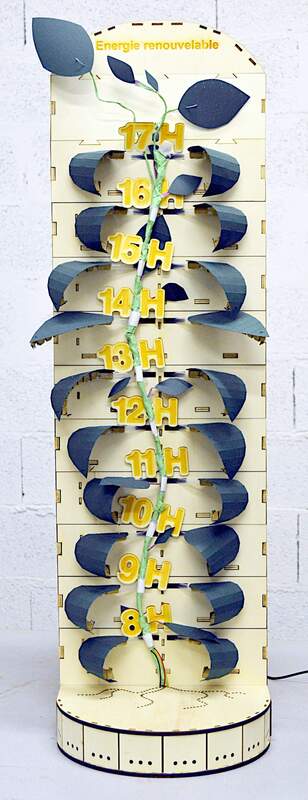} \caption{From 12:00 to 15:59, peak at 14H.}
    \label{fig:teaser:var4}
  \end{subfigure}\hspace{\spaceW}
  \begin{subfigure}[b]{\figWratio\linewidth}
    \centering
    \includegraphics[height=\protoH]{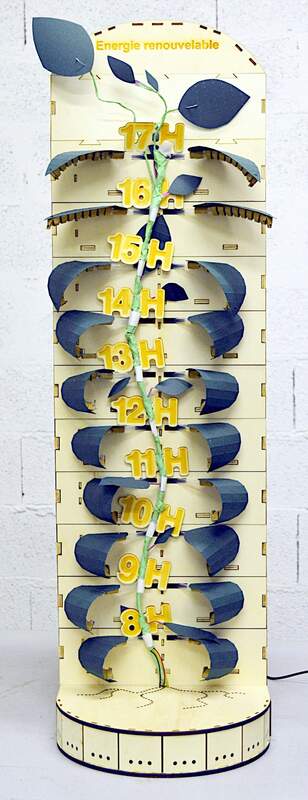} \caption{From 15:00 to 17:59, peak at 16H.}
    \label{fig:teaser:var5}
  \end{subfigure}
  \caption{Five renewable-energy variations conveyed with PlantFORM. This plant-like shape-changing interface, which encodes data series through folded shapes, serves as a physical vertical chart.}
  \label{fig:teaser}
  \Description{The five pictures are photographs of the PlantFORM prototype, which stands as a wooden structure with a horizontal basis, and a vertical rear wall. The rear wall supports the trunk of the plant-like chart. The leaves are unfurled according to the displayed energy rates. Laser-cut wooden labels are placed upon each pair of leaves to show the corresponding hour.}
\end{teaserfigure}
 
\maketitle

\def\Tdischarge{$T\ts{\textit{discharge}}$}
\def\Trecharge{$T\ts{\textit{recharge}}$}

\section{Introduction}\label{sec:introduction}

Conveying scientific data to non-expert readers has grown interest in environmental concerns, as well as data-driven strategies that help users reduce their environmental impact (e.g., Eco-feedback interfaces \cite{chalal_visualisation_2022,froehlich2010design}, Sustainable HCI (SHCI) \cite{disalvo2010mapping,silberman2014next,bremer_have_2022}).
However, much of the environment-related scientific data evolves gradually and is provided by energy management interfaces for the following hours \cite{daniel2019cairnform,schrammel_forewatch_2011,kjeldskov_eco_forecasting_2015}, days \cite{costanza_doing_laundry_2014,bourgeois_conversations_2014}, or both \cite{simm_tiree_2015}, thus involving daily use.
Therefore, conveying data series on environmental concerns must benefit from visualizations that better integrate into daily living environments than rigorous, neutral, and stark scientific visualizations.

Previous work has already studied the advantages of embellished charts and infographics and showed advantages in memorability \cite{borkin2013memorable,haroz2015visualization}, more careful reading \cite{andry_interpreting_2021,haroz2015visualization}, and a feeling of lower effort \cite{andry_interpreting_2021}, which fit well scopes out of science and engineering, such as popular magazines or journals. Moreover, embellishment only marginally decreases understandability \cite{andry_interpreting_2021}, even if traditional infographics and charts remain slightly more accurate \cite{bateman2010junk,skau2015evaluation} and clear \cite{andry_interpreting_2021,bateman2010junk}.
Furthermore, nature is already an inspiration source for several environmental visualizations (e.g., trees \cite{coutaz_will_2018,dimitriou_charged_2018,piccolo_socially_2017,vilarinho_combining_2016}, flowers \cite{backlund2007static,rist_next-generation_2012}, and plants \cite{plichta2023growing,holstius2004infotropism}), mainly because the link to nature reminds users that they act for the environment \cite{chalal_visualisation_2022}.
Consequently, this paper addresses the following questions: Can we encode data series with plants? Can we implement plant-like charts combining effective reading and high aesthetics?

Prior research does not provide answers to these questions.
Bio-inspiration is an active and prolific field that has provided solutions in innovation, technology, and engineering in robotics \cite{xu2021recent} and soft robotics \cite{meder2023perspective} over the past years by mimicking plants' growth, adaptability, materials, sustainability, and effectiveness.
Such inspiration from nature also provided new designs to convey digital information.
For instance, some recent work on shape-changing interfaces animate natural plants to
convey a gradient of air quality (from good to medium to bad) \cite{seow_pudica_2022},
materialize a two-state gauge between trash and recycling behaviors \cite{holstius2004infotropism},
or maps a daily step counter \cite{chien2015biogotchi},
and create artificial vegetation that
materializes a rate of households heating accomplishment \cite{plichta2023growing},
requests attention through motion for health habits (from subtle to heavy notification) \cite{lee2023decorative},
or, augments viewing experience as ambient effects for video games \cite{degraen2020ambiplant}.
However, data series remain to explore.

This paper investigates these questions by following a research-through-design approach \cite{zimmerman2007research,zimmerman2010analysis} because of research outcomes that include the design process to get the right user interface (i.e., preferred state) for a specific framing within the real world.
Our real-world goal is to design a shape-changing artifact conveying daily renewable-energy forecasts at workplaces for ten-hour working days in semi-public places like shared offices. This artifact must enable users to retrieve information about renewable-energy availability, such as energy variations' peaks and slopes.

Consequently, this research investigates the readability and aesthetics of a plant-like shape-changing physical interface displaying a ten-hour data series of energy rates. Initially, we define and explore a design space along four axes: trunk form, anchoring, decoration, and animation.
The exploration of this design space leads to comparing eight low-fidelity histograms through two online studies (N=16 and N=25) that inspected three aspects of interaction with embellished charts: readability, clarity, and aesthetics.
Subsequently, we examine some implementation aspects, like modality and materials, which are tested through a comparative user study (N=28) with four high-fidelity prototypes.
During those three studies, the participants answered questions about storage use involving peak and slope retrieval tasks.
Finally, some insights are provided on the learnability of rates encoding by folded shapes.

The studies bring three main results.
First, getting naturalness and keeping readability compel compromise that restricts the design possibilities of plant-like charts.
Whereas curvy and straight trunks reach similar readability rates, alternated anchorings bring higher aesthetics than one-sided anchorings but lead to low readability and clarity. 
Second, the choice of materials influences tech trend perception and naturalness.
Whereas wood, cardboard, and thick paper materials maximize the naturalness of a physical plant-like chart, a physical ring chart using illuminated PMMA-made rings\footnote{PMMA is a translucent plastic material that ressembles glass by aspect, resistance, and optical properties but with light weight (i.e., polymethyl methacrylate, a.k.a., Plexiglass $\circledR$). This material is suitable for laser cutting fabrication.} is perceived as the most innovative.
Third, rate encoding using folded shapes is understandable without explanations if encoding rates from 0 to a maximum value encoded by the horizontal line position.
Therefore, the studies validate the design choice of a two-sided anchoring version of a plant-like chart with a curvy trunk that uses unfurling leaves to encode rates.

In summary, the four main contributions of this paper are:
\begin{enumerate}
\item A design space for plant-like vertical charts;
\item The design and implementation of a shape-changing plant-like artifact that maximizes readability and aesthetics to convey data series; \item Results on the effect of material choice on tech trend perception and user preferences;
\item Results on readability and learnability of folding shapes to encode rates.
\end{enumerate}

This paper first reviews the related work and then motivates the design intents that guided the research-through-design approach of this work.
Then, we introduce a design space of plant-like vertical charts along four dimensions, which we explore through two online studies that lead to one possible design maximizing readability and aesthetics.
After describing our implementation of a physical plant-like chart, which uses a folding mechanism, we present a user study that compared this prototype to three others through varying display modalities and materials choices.
We conclude by discussing the findings and highlighting the limitations of the work.

\section{Related Work}\label{sec:relatedwork}

This section reviews related work on designing eco-forecast interfaces for shifting energy consumption, using nature as a metaphor for environmental data, and studying charts' embellishment and physicality.

\subsection{Eco-Feedback and Eco-Forecasts}

An HCI approach to sustainability is to provide users with feedback \cite{froehlich2010design} on their resources' consumption \cite{coutaz_will_2018}, savings \cite{dimitriou_charged_2018,piccolo_socially_2017}, or waste \cite{kim_coralog_2009} or about the environmental consequences of their lifestyle \cite{stegers_ecorbis_2022}. These Eco-Feedback interfaces make consequences visible so users can react by anticipation (principle of self-monitoring \cite[Ch.~3]{fogg2003persuasive}).

Another approach is to make resource status visible so that users can immediately adapt or plan their resource usage. For example, some physical ambient interfaces display the current status of energy production \cite{pierce_local_2012,pierce_materializing_2010,pierce_beyond_2012,quintal_wattom_2019,quintal_watt-i-see_2016} or energy storage \cite{elbanhawy_towards_2016}. Another example is to display the upcoming status of resources through Eco-Forecast interfaces \cite{kjeldskov_eco_forecasting_2015} so users can shift energy usage (e.g., at home \cite{bourgeois_conversations_2014,jensen_washing_2018,kjeldskov_facilitating_2015,schrammel_forewatch_2011,simm_tiree_2015}) or consumption (e.g., at the workplace \cite{daniel2021cairnform}). Shifting consumption requires storage capacities to supply appliances when renewable energy is low or unavailable, thus keeping appliance usage unaltered. From then on, users do not plan appliance usage; instead, they plan storage use and recharge. Those strategies are to become widely possible using electric vehicles' batteries as grid support \cite{brush_evhomeshifter_2015,tomic_fleets_2007}. Therefore, this research targets the consumption shift using storage at office workplaces, where workers' energy usage is usually tricky to postpone.

\subsection{Nature as Metaphor of Energy Data}

The literature on Sustainable HCI regularly associates energy data with environmental symbolism, such as nature, landscapes, or trees. Almost all are graphical representations. For example, some representations involve wild animals \cite{dillahunt_understanding_2014,kim_coralog_2009} and natural landscapes or gardens \cite{nisi_sinais_2013,pereira_understanding_2020,pittarello_public_2019,rist_next-generation_2012}. Some others are trees that grow based on energy savings and challenge achievement \cite{dimitriou_charged_2018}, fill according to energy scores \cite{vilarinho_combining_2016}, or change color, height, and fullness according to teams' consumption and virtue \cite{coutaz_will_2018}. Combining visualizations of environmental data with such artistic representations tends to increase concern about the environment of users who feel affection for those interfaces \cite{chalal_visualisation_2022}. However, even if already deeply explored in the literature, augmented plants and plant-like artifacts include only a few energy metaphors.

\subsection{Shape-Changing Plants}

For two decades, the exploration of Human--Plant Interaction (HPI) produced a variety of plant or plant-like prototypes \cite{chang_patterns_2022,loh2024more-than-human,webber2023engaging}. Shape change and actuation are among the possible output modalities with natural or artificial plants \cite{chang_patterns_2022}. Examples comprise orientating natural stems passively with lightening \cite{holstius2004infotropism} or opening and closing natural leaves through air stimuli \cite{gentile2018plantxel} or electrostimulation \cite{seow_pudica_2022}. Some other examples combine the orientation and shaking of artificial plants \cite{degraen2019overgrown,degraen2020ambiplant,hong2015better,lee2023decorative} and animate plant elements by growing blades of grass \cite{tanaka2024programmablegrass} or movable petals or stems \cite{antifakos2003laughinglily,chooi2022symbiosis,degraen2022familyflower,plichta2023growing}. Finally, one example animates plants by furling some leaves through shape memory alloy in response to touch \cite{cheng2014mood-fern}.

Human--Plant Interaction encompasses various applications \cite{loh2024more-than-human} ranging from learning or caring about plants to shaping health or recycling behaviors to sustaining resources such as air quality or energy. However, displaying energy data through plants remains underexplored. The three examples retrieved include a flower artifact for heating energy consumption \cite{plichta2023growing}, an energy-saving plant that is watered to promote stairs instead of a lift \cite{loh2010please}, and a tree-like artifact for energy savings in schools \cite{piccolo_socially_2017}. This latter example, SEETree \cite{piccolo_socially_2017}, is a physical tree representing daily energy-saving scores through seven illuminated branches. This illuminated tree was used in an elementary school---a semi-public space---to stimulate group involvement. Sustaining this inspiring exploration effort is worth it, as nature can inspire shape change \cite{qamar_morphino_2020,yao2024info-motion}.

Furthermore, plants are already used to represent or convey various kinds of data: for example, a rate \cite{plichta2023growing}, a counter \cite{chien2015biogotchi,degraen2019overgrown}, continuous or discrete gauges between some states \cite{antifakos2003laughinglily,holstius2004infotropism,hong2015better,seow_pudica_2022}, bitmap displays \cite{gentile2018plantxel,takaki2014mossxels,tanaka2024programmablegrass}, ambient effects \cite{degraen2020ambiplant}, emotions \cite{angelini2016multisensory,bhat2021plant-robot,cho2015emotional}, notifications \cite{hong2015better,lee2023decorative}, or awareness \cite{degraen2022familyflower}, but data series are still underexplored.
Therefore, encoding energy data series with plants is still to be explored in depth.

\subsection{Charts' Embellishment and Physicality}

Embellishment changes user perception of data from purely textual and analytical interpretation to visual communication. Previous research studied embellished graphical infographics from communication media handmade or customized by artists or graphic designers \cite{andry_interpreting_2021,arunkumar2024image,bateman2010junk,borkin2013memorable,burns2022pictographs} and the embellishment of computer-generated bar charts \cite{haroz2015visualization,skau2015evaluation,skau2017readability}. Although the reading accuracy of infographics from communication media is no worse than plain charts \cite{bateman2010junk}, more systematic investigations revealed that bar shapes can highly impact error rates \cite{skau2015evaluation}, whereas pictorial bars have no discernible impact \cite{skau2017readability}.
However, inspecting embellished charts also better engages users in more careful reading \cite{andry_interpreting_2021,haroz2015visualization}, whereas the speed of finding information remains similar when only pictorials are added to plain bars \cite{burns2022pictographs}. Nevertheless, extra pictographs are harmful if out of data mapping \cite{haroz2015visualization}.

Preserving the presence of chart elements (such as axes, labels, or captions) is required to perceive these infographics as a source of information rather than pure imagery \cite{arunkumar2024image}. Moreover, the general preference for embellished infographics is only moderated because plain charts are also liked \cite{andry_interpreting_2021,bateman2010junk,burns2022pictographs}. Indeed, even if embellishment is generally found to be easiest \cite{bateman2010junk,burns2022pictographs}, users also feel that plain charts are clearer and faster to read \cite{andry_interpreting_2021,bateman2010junk}. An explanation for embellished charts being more liked is that they are more aesthetically pleasing \cite{andry_interpreting_2021}. Aesthetics then ought to be a decisive design factor for successful embellishments of charts, as well as the presence of chart elements to evoke an information source and the choice of bar shapes enabling enough accuracy. Also, as these previous works on embellishment focused on understanding static graphical charts, studying dynamic physical charts remains original.

Some work studied physical shape-changing charts of the traditional form (i.e., not specifically designed for a particular data type, such as inFORM \cite{follmer_inform_2013} and EMERGE \cite{taher_exploring_2015}).
Studies show that physical charts are more effective than graphical charts based on low-level information retrieval tasks (i.e., range, order, and comparison) \cite{jansen_evaluating_2013}.
Further results indicate that physical charts provide better recall of quantitative data about countries \cite{stusak_evaluating_2015} and improve data memorability when the data set is also interesting and comprehensible \cite{stusak_if_2016} compared to their graphical versions.
Physical charts also responded considerably faster than in virtual reality environments \cite{ren_comparing_2021} when asking high-level questions on non-trivial data sets.

Physical charts have been little explored to present energy data. One exception is CairnFORM \cite{daniel2019cairnform}, a vertical histogram with dynamic rings displaying daily renewable energy forecasts. Whereas CairnFORM is effective for comparison and range tasks, this histogram is only marginally efficient for the scheduling task. Some preliminary results also indicate a tendency for the hedonic and pragmatic qualities of CairnFORM to increase over two months; in contrast, these qualities decreased for the graphical version of the histogram \cite{daniel2021cairnform}. This tendency encourages using physical charts for a better daily user experience. We propose to design and study an embellished version of this vertical shape-changing histogram.

Before introducing and exploring a design space for plant-like charts and implementing and evaluating a prototype, the following section presents the design intents that motivate nature-evocative charts.

\section{Design Intents: Convey Energy Data with Nature-Evocative Charts}\label{sec:intents}

Regarding the research-through-design approach \cite{zimmerman2007research,zimmerman2010analysis} that we follow, this section summarizes the design intents that led to a plant-like shape-changing chart to convey renewable energy forecasts in public spaces.

\def\footnotevariation{\footnote{An energy variation starts from a minimum energy availability (local minimum or global minimum), increases until a maximum (local maximum or global maximum), and decreases until the following minimum (local minimum or global minimum) \cite{daniel2019cairnform}.}}

\subsection{Shared Display at Workplaces}

The Sustainable HCI community now focuses on groups or communities instead of individuals \cite{bremer_have_2022}. This way, similarly to Watt-Lite \cite{katzeff_exploring_2013} and CairnFORM \cite{daniel_shape-changing_2021}, we target workplaces because managing the hustle and bustle of daily life in households through rational indicators has shown several limitations \cite{dourish2010politics,strengers2011designing,strengers2014resource}. In contrast, working hours are more structured and routine. Moreover, shared offices allow for sharing a single display all day long per user group. Furthermore, workplaces are suitable places for microgeneration\footnote{Microgeneration is small-scale electricity or heat generation by individuals, businesses, or communities to meet their own needs.}, where storage management by users could help shift consumption without shifting usage if provided with renewable energy forecasts. A previous study showed that several employees in shared offices could use one display over several weeks \cite{daniel_shape-changing_2021}.

\subsection{Physical Embodiment of Data Series}

We propose to convey forecasts for ten-hour working days (from 8:00 am to 5:59 pm) through ten-value data series of mean renewable energy rates per hour. However, renewable energy variations\footnotevariation{} are to be displayed one after another because they are easier to read than when displayed all at once \cite{daniel2019cairnform}.

This way, we favor a physical embodiment of energy forecasts because the physical presence among the users is an enabler for discussion \cite{sauve_econundrum_2020}. This physical display is to be placed in workplaces where every worker can easily observe it. Therefore, renewable-energy forecasts are available to anyone at any time, who can then be informed of the opportunities for storage use and recharge. However, we prefer using shape-changing artifacts, such as CairnFORM \cite{daniel2019cairnform}, rather than a static one, such as SEETree \cite{piccolo_socially_2017}.

When updating to show the following variation, physical shape-change will play as an ambient notification, which we expect to be more pleasing and fascinating than a color change. When placed ideally, we expect workers to perceive motions of our display with their peripheral vision when their central attention and vision are already focused on a primary task. Indeed, systems leveraging peripheral interactions allow users to multitask and interact spontaneously \cite{sauve_econundrum_2020,stegers_ecorbis_2022}. Moreover, some speed motions of shape-changing interfaces can notify users without disturbing them \cite{daniel2019cairnform}.

\subsection{Nature as Design Inspiration}

The appearance of our display must inform about the purpose it supports, as well as the kind of data it presents. We favor an artistic representation of data \cite{chalal_visualisation_2022} that arouses curiosity, discussion, and contemplation so that our display could also stimulate collective reflections on renewable energy. Furthermore, combining data visualizations for sustainability with nature-inspired artistic representations is recommended because users \textit{``who develop an emotional attachment to nature-inspired representations show more concern for the environment''} \cite{chalal_visualisation_2022}.

Following the example of ChArGED \cite{dimitriou_charged_2018} or CoSSMunity \cite{vilarinho_combining_2016}---which resort to tree-like data representations---we chose plant-like charts to display 10-hour forecasts of renewable energy rates with an emotional visual inceptive in the form of a living plant. Instead of classical bars, this embellished chart encodes rates through nature-inspired shapes, which are leaves animated through wave-like furling-unfurling motion \cite{qamar_morphino_2020} (see \cref{fig:design_inspiration:bio1,fig:design_inspiration:bio2,fig:design_inspiration:bio3,fig:design_inspiration:bio4}). The more a leaf is unfurled, the greater the availability rate of renewable energy (see \autoref{fig:design_inspiration:ornementation}). The following section proposes a design space for such plant-like charts.

\def\factorA{0.5}
\def\factorB{0.135}
\def\height{1.8cm}
\newcommand{\license}{
{\tiny
  Licenses:
  (a) Geoff McKay, \href{https://creativecommons.org/licenses/by/2.0/}{CC BY 2.0};
  (b) Dominicus Johannes Bergsma, \href{https://creativecommons.org/licenses/by-sa/4.0/}{CC BY-SA 4.0};
  (c) \href{https://creativecommons.org/publicdomain/zero/1.0/}{CC0};
  (d) Aditya Gurav, \href{https://creativecommons.org/licenses/by-sa/4.0/}{CC BY-SA 4.0}.
  }
}
\begin{figure}
\begin{subfigure}[b]{0.11\linewidth}
    \centering
    \includegraphics[height=2cm]{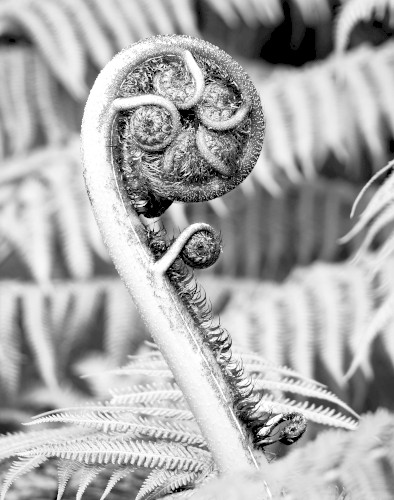}
    \caption{Fern koru.}
    \label{fig:design_inspiration:bio1}
    \Description{The picture is a photograph of a ready-to-unfurl fern koru. The extremity of the central branch is furled, with smaller internal structures that are also furled.}
  \end{subfigure}\begin{subfigure}[b]{0.135\linewidth}
    \centering
    \includegraphics[height=2cm]{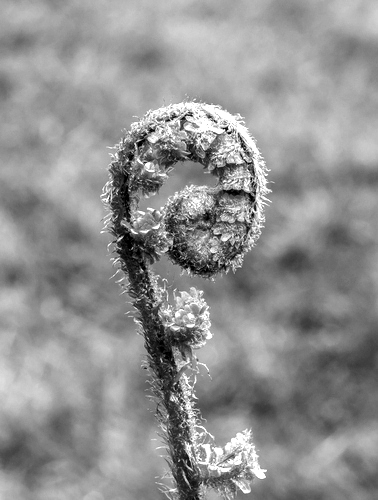}
    \caption{Young fern.}
    \label{fig:design_inspiration:bio2}
    \Description{The picture is a photograph of a furled young fern. The extremity of the young furn is furled.}
  \end{subfigure}\begin{subfigure}[b]{0.135\linewidth}
    \centering
    \includegraphics[height=2cm]{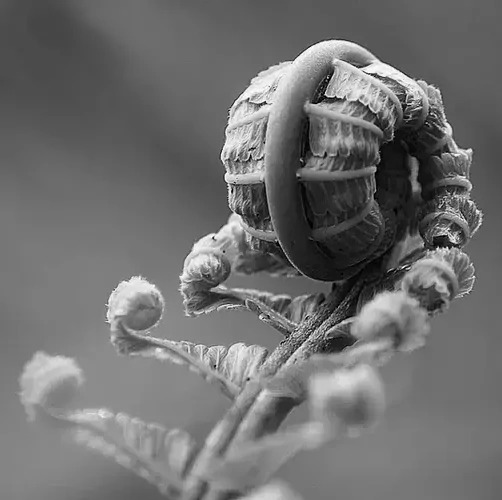}
    \caption{Fern leaf.}
    \label{fig:design_inspiration:bio3}
    \Description{The picture is a photograph of a fern leaf whose extremity is furled. Along the central stem that is furled, the lateral elements of the leaf, anchored along the central stem, are also furled.}
  \end{subfigure}\begin{subfigure}[b]{0.150\linewidth}
    \centering
    \includegraphics[height=2cm]{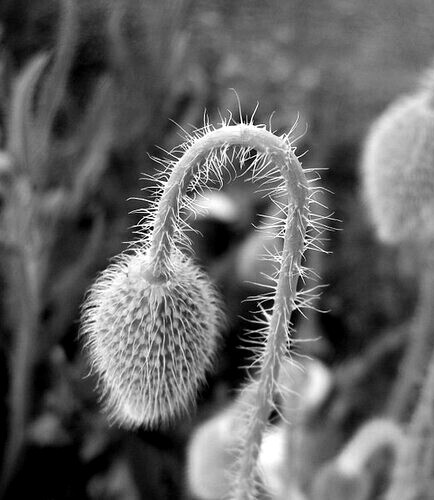}
    \caption{Poppy bud.}
    \label{fig:design_inspiration:bio4}
    \Description{The picture is a photograph of a poppy bud that is closed. The poppy bud is oriented down because it is hung by a folded trunk.}
  \end{subfigure}
 \begin{subfigure}[b]{0.46\linewidth}
    \centering
    \includegraphics[width=0.9\linewidth]{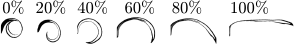}
    \caption{Encoding rates by leaf unfurling.}
    \label{fig:design_inspiration:ornementation}
    \Description{The drawing of a plant leaf as histogram ornamentation that unfolds through six positions from 0\% to 20\%, to 40\%, to 60\%, to 80\%, and to 100\%.}
  \end{subfigure}
  \license{}
  \caption{
    Plants' furling inspired our design of rate encoding.
    The design intent is to relate energy rates to nature's concerns.
}
  \label{fig:design_inspiration}
\end{figure}
 
\section{Framing: A Design Space for Plant-Like Vertical Charts}\label{sec:design-space}

\autoref{fig:design-space} provides a design space for plant-like vertical charts to rationalize and systematize the exploration of possible designs.
The values of each axis are sorted from lower to higher nature-likelihood.
The four axes are the following:

\def\factor{0.5}
\begin{figure}
  \includegraphics[width=\factor\linewidth]{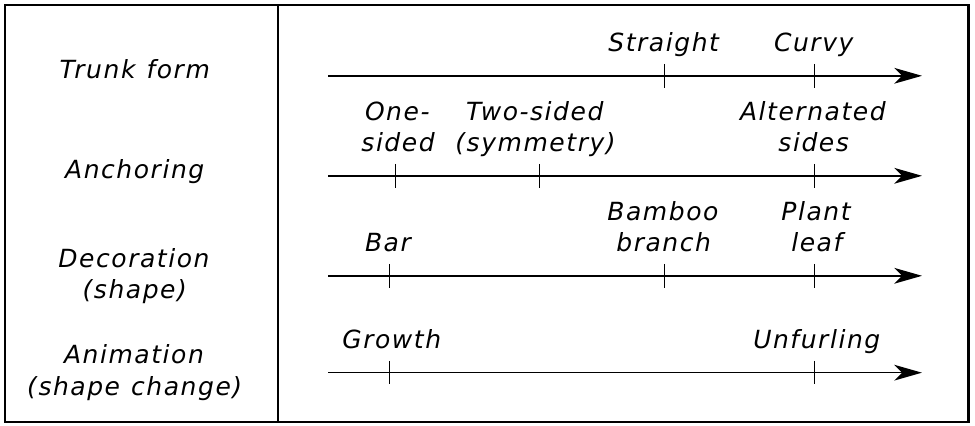}
  \caption{
    A four-axis design space for plant-like vertical charts.
    The values are sorted from lower to higher nature-likelihood.
  }
  \label{fig:design-space}
  \Description{The names of the four axes are listed in a column at the left: Trunk form, Anchoring, Decoration (shape), and Animation (shape change). On the right, four oriented axes show the values of each axis. The values are placed on the axes at distances reflecting their naturalness.}
\end{figure}

\begin{enumerate}
\item \textit{Trunk form}: What trajectory does the trunk follow? \item \textit{Anchoring}: Where are the decorations anchored on the trunk?
\item \textit{Decoration}: What kind of shape is encoding rates? \item \textit{Animation}: How are shapes updated when the rates change? \end{enumerate}

The first axis is the form of the trunk, the central structure supporting all other aerial structures of a plant.
Both \textit{straight} trunks and \textit{curvy} trunks exist in nature.
However, curvy trunks may be perceived as more natural than straight ones in the representation of nature that resides in peoples' minds.

The second axis, called anchoring, is the global pattern followed when connecting other aerial structures of a plant on the central trunk.
Anchoring is \textit{one-sided} when all structures are connected on the same trunk side.
This anchoring is the closest to classical histograms but is rare in nature.
A \textit{two-sided} anchoring is when the connected structures of a one-sided anchoring are symmetrically duplicated on the other side of the trunk.
Again, this anchoring is rare in nature.
An anchoring on \textit{alternated sides} is when aerial structures are connected on each side of the trunk every other time.
Even if such regularity is also rare in nature, this anchoring may look the most natural to users.

The third axis, on decorations, describes the aerial structures connected to the plant's trunk.
The three values are \textit{bars}, \textit{bamboo branches}, and \textit{plant leaves}.
Bars differ from naturelike shapes, but using the highest values of the other axes can be sufficient to create nature-evocative charts (e.g., a curvy trunk and an alternated anchoring).
Even if bamboo branches are a biological aberration, because bamboo plants grow as a single branch that is their trunk, those straight shapes are close to the bars found in classical histograms.

Finally, the way the decorations are animated to encode data stands as the fourth axis.
Indeed, decorations must encode data from a minimum to a maximum value.
The first possible animation is \textit{growth}.
As in biology, decorations start from an early development state and grow their shape until a more mature development state.
However, growth animations bear little biological resemblance because displaying data that evolves implies reversing growth at some times, which is not natural.
The second animation is \textit{unfurling}.
Decorations encode the minimum value through the folded state, then unfurl until the unfolded state, which encodes the maximum value.
Furling and unfurling motion are found in nature, for example, when flowers unfold their petals in the morning or when plants' aerial structures follow such motion cycles according to water inflows.
Moreover, furling of branches and trunks also exists in nature (see \cref{fig:design_inspiration:bio1,fig:design_inspiration:bio2,fig:design_inspiration:bio3,fig:design_inspiration:bio4}).

We aim to design a nature-like display evoking plants to users at best (i.e., strictly respecting all biology's rules is neither intended nor desired).
Even if evoking nature is required, our display must remain readable to users.
Consequently, the following two sections explore this design space to get a plant-like chart matching the criteria of readability and aesthetics.
Eight design choices are compared through two online studies using low-fidelity graphical histograms (to save time and budget).
Further physical prototyping iterations happened only on the final histogram choice.

\section{First Online Study: Exploring Trunk, Anchoring, and Decoration Effects}\label{sec:online1}

This first online study (N=16) aims to understand better the effects of trunk, anchoring, and decoration choices on chart readability and user preferences for clarity, interest, and aesthetics.
Six designs are involved on a continuum from more classical to more nature-like histograms (see \autoref{fig:online1:histograms}).
The more classical one is a straight, one-sided bar-like chart, and the more nature-like one is a curvy alternated-leaves chart.
The four intermediate versions vary independently in trunk, anchoring, and decoration to better understand the effect of each factor.

\def\factor{0.4}
\def\height{5cm}
\begin{figure}
  \begin{subfigure}[b]{\factor\linewidth}
    \centering
    \includegraphics[height=\height]{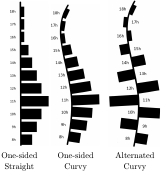}
    \caption{Three vertical bar charts.}
    \label{fig:online1:histograms:bars}
    \Description{The three bar histograms of the first online study display an eleven-hour energy variation used during the first online study.}
  \end{subfigure}
  \begin{subfigure}[b]{\factor\linewidth}
    \centering
    \includegraphics[height=\height]{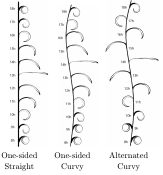}
    \caption{Three vertical leaf charts.}
    \label{fig:online1:histograms:leaves}
    \Description{The three leaf histograms of the first online study display an eleven-hour energy variation used during the first online study.}
  \end{subfigure}
  \caption{
    First Online Study -- The six low-fidelity histograms.
    Four dimensions are explored: trunk form (straight vs curvy), anchoring (one-sided vs alternated), decoration (bar-like vs leaf), and animation (growth vs unfurling).
    The histograms are ordered on a continuum of more classical to more nature-like histograms, from left to right.
  }
  \label{fig:online1:histograms}
\end{figure}
 
\subsection{Protocol}

The six histograms were sketched in non-animated image files and included in online questionnaires\footnote{Google Forms is an online survey web application (\href{https://docs.google.com/forms/}{https://docs.google.com/forms/}).}.
Renewable energy rates ranged from 0\% to 100\%, with the degree of leaf unfolding corresponding to the magnitude of the renewable energy rate (see \autoref{fig:design_inspiration:ornementation}).
For each presented data series, participants were asked to identify peak hours and hour ranges of ascending or descending energy production.
Therefore, tasks involved reading and interpreting one renewable energy variation per histogram (see \autoref{tab:online1:variations}) to optimize storage or recharge through scenarios involving the daily management of a laptop battery.

\begin{table}
  \centering {\tablefontsize
  \caption{
    First Online Study -- Design of the six variations assigned to the six histograms.
    Those variations are inspired from real solar-panel production data registered at our lab.
  }~\label{tab:online1:variations}
  \begin{tabular}{lllrrrrclllrrrr}
\cmidrule[0.5pt]{1-7}\cmidrule[0.5pt]{9-15}
\it Shape &\it Sides &\it Trunk  &\it Length &\it Start &\it Peak &\it End && \it Shape &\it Sides &\it Trunk  &\it Length &\it Start &\it Peak &\it End\\
\cmidrule{1-7}\cmidrule{9-15}
\ifHOURA
  Bar  & One  & Straight &  8h &  8:00 & 11:00 & 15:59 &&  Leaf & One  & Straight & 10h &  9:00 & 13:00 & 17:59\\
  Bar  & One  & Curvy    &  8h &  8:00 & 11:00 & 15:59 &&  Leaf & One  & Curvy    &  9h & 10:00 & 14:00 & 18:59\\
  Bar  & Alt. & Curvy    &  8h &  8:00 & 11:00 & 15:59 &&  Leaf & Alt. & Curvy    &  9h & 10:00 & 14:00 & 18:59\\
\else
  Bar  & One  & Straight &  8h &  8H & 11H & 15H       && Leaf & One  & Straight & 10h &  9H & 13H & 17H \\
  Bar  & One  & Curvy    &  8h &  8H & 11H & 15H       && Leaf & One  & Curvy    &  9h & 10H & 14H & 18H \\
  Bar  & Alt. & Curvy    &  8h &  8H & 11H & 15H       && Leaf & Alt. & Curvy    &  9h & 10H & 14H & 18H \\
\fi
\cmidrule[0.5pt]{1-7}\cmidrule[0.5pt]{9-15}
  \end{tabular}
  }
\end{table}
 
Sixteen participants (5 females and 11 males, from 26 to 63 years old, 38.6 years old on average, SD=11.4) were recruited by sending e-mails to university departments, campus companies, and academic mailing lists and were assigned to three groups for this within-subject study.
Histograms' presentation was counterbalanced through three condition orders by following Latin squares.
Based on previous work on the embellishment of charts and infographics \cite{andry_interpreting_2021,bateman2010junk}, perceived aesthetics, interest, and clarity were rated through five-point Likert scales.
In addition, the participants were asked to choose the histogram they preferred to perform the tasks.

\subsection{Assumptions}

We draw four assumptions on the effect of trunk, anchoring, and decoration factors on readability and user preferences:

\begin{itemize}

\item[(A1)] \textbf{Leaf charts are more complicated to read [+10\%, +20\%] than bar-like ones} because people are less used to them and because decoding folded shapes is harder than bars, particularly to discriminate the zero position that not only requires length analysis but shape analysis, thus requiring more cognition and making the reading of variations' slopes harder.

\item[(A2)] \textbf{Histograms with curvy trunks are slightly more complicated to read [+5\%, +10\%] than ones with straight trunks} because the one-to-one comparison of shapes is only slightly altered, and the visual reconstruction of the slope remains easy and fast thanks to visual persistence.

\item[(A3)] \textbf{Alternated anchoring is harder to read [+40\%, +60\%] than one-sided anchoring} because reading the data slope by one-to-one comparison of shapes cannot rely on visual persistence and must resort to cognition and memory to match the shapes from each side.

\item[(A4)] \textbf{The more plant-like, the most preferred aesthetics} because embellishment is more unusual, less neutral, and has better evocation of renewable energy data.

\end{itemize}

\subsection{Results and Analysis}

\begin{figure*}
  \begin{subfigure}[b]{0.33\linewidth} \centering
    \includegraphics[width=1.0\linewidth]{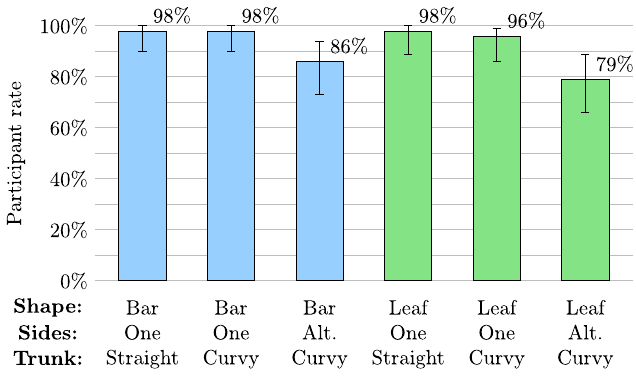}
    \caption{Task success.}
    \label{fig:results:online1:success}
    \Description{The bar chart shows the results of task success for the six histograms of the first online study.}
  \end{subfigure}
\begin{subfigure}[b]{0.33\linewidth}
    \centering
    \includegraphics[width=1.0\linewidth]{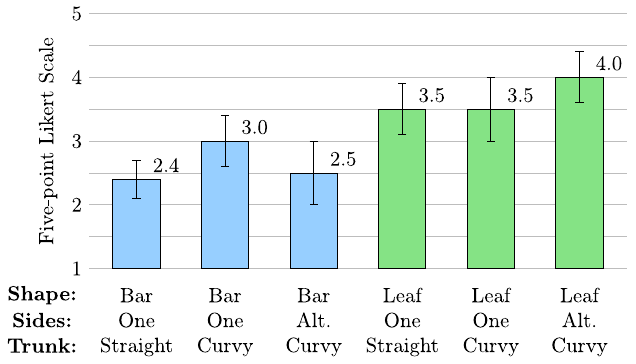}
    \caption{Preference: Aesthetics.}
    \label{fig:results:online1:aesthetics}
    \Description{The bar chart shows the results on aesthetics for the six histograms of the first online study.}
  \end{subfigure}
\begin{subfigure}[b]{0.33\linewidth}
    \centering
    \includegraphics[width=1.0\linewidth]{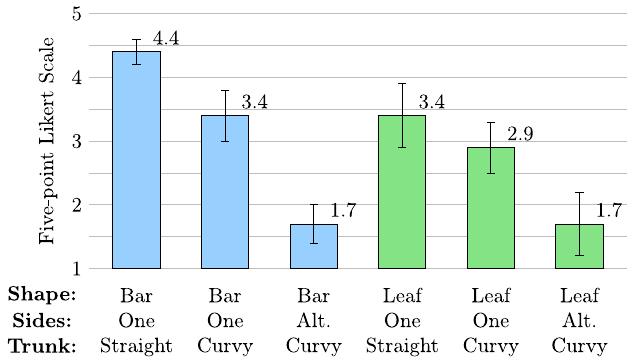}
    \caption{Preference: Clarity.}
    \label{fig:results:online1:clarity}
    \Description{The bar chart shows the results on clarity for the six histograms of the first online study.}
  \end{subfigure}
    {\parbox{0.95\linewidth}{\scriptsize
      \textit{Note.} \textbf{Error bars} are 95\% confidence intervals from Jeffreys Bayesian method for success rates, from a normal distribution for normally distributed scores, and from the percentile bootstrap method for non-normally distributed scores.
    }}
  \caption{First Online Study -- Task success and preferences on aesthetics and clarity for the six histograms (N=16).}
  \label{fig:results:online1}
\end{figure*}

The results in \autoref{tab:results:online1} and \autoref{fig:results:online1} align with previous results regarding embellished infographics' preferred aesthetics \cite{andry_interpreting_2021} but lower clarity \cite{andry_interpreting_2021,bateman2010junk} and decreased accuracy \cite{skau2015evaluation}.
Indeed, despite the alternated leaves with a curvy trunk version having the closest design to nature and getting the highest aesthetics, the traditional straight histogram with one-sided bars was rated as the most interesting and clear and is the most preferred one to perform the task.
The results also reveal that alternated anchoring was the most challenging to read, the least clear, and the least preferred.
Trunk form had only a marginal impact on the high success rate of one-sided anchorings.

\begin{table*}
  \caption{First Online Study -- Results for the six histograms (N=16).}
  \label{tab:results:online1}
  {\tablefontsize
    \ifPERA
    \begin{tabular}{
        p{15pt}p{15pt}p{30pt}
        @{}p{1pt}
        p{38pt}@{\hspace{8pt}}p{44pt}@{\hspace{8pt}}p{44pt}@{\hspace{8pt}}p{40pt}
        @{}p{1pt}
        p{31pt}@{\hspace{8pt}}p{31pt}@{\hspace{8pt}}p{31pt}@{\hspace{8pt}}p{31pt}
        @{\hspace{8pt}}p{38pt}
      }
    \else
    \begin{tabular}{
        p{15pt}p{15pt}p{25pt}
        @{}p{1pt}
        p{23pt}@{\hspace{8pt}}p{28pt}@{\hspace{8pt}}p{28pt}@{\hspace{8pt}}p{44pt}
        @{}p{1pt}
        p{33pt}@{\hspace{8pt}}p{33pt}@{\hspace{8pt}}p{33pt}@{\hspace{8pt}}p{33pt}
        @{\hspace{8pt}}p{38pt}        
      }
    \fi
    \midrule
    \multicolumn{3}{l}{\tbh{Histogram}} && \multicolumn{4}{l}{\tbh{Success Rates}} &&  \multicolumn{5}{l}{\tbh{Preferences}}\\
    \cmidrule{1-3}\cmidrule{5-8}\cmidrule{10-14}
    \textit{Shape}&\textit{Sides}&\textit{Trunk} && \textit{T\textsubscript{peak}} & \textit{T\textsubscript{ascending}} & \textit{T\textsubscript{descending}} & \textit{Overall} && \textit{Aesthetics} & \textit{Interest} & \textit{Clarity} & \textit{Overall} & \textit{Preferred} \\
    \midrule
\ifPERA
    Bar  & One  & Straight  & &\raL\bf\hfill88\%~{\tiny [66\%,~97\%]} &\raL\bf\hfill94\%~{\tiny [74\%,~99\%]} &\raL\bf\hfill100\%~{\tiny [86\%,~100\%]} &\raL\bf\hfill94\%~{\tiny [84\%,~98\%]}& &\raL   {\tiny$^{\vartriangle}_{\bullet}$}\hfill2.4$\pm$0.3 &\raL\bf{\tiny$^{\vartriangle}_{\bullet}$}\hfill3.7$\pm$0.4 &\raL\bf{\tiny$^{\vartriangle}_{\bullet}$}\hfill4.4$\pm$0.2 &\raL\bf{\tiny$^{\vartriangle}_{\bullet}$}\hfill3.5$\pm$0.2 &\raL\bf 38\%~{\tiny [17\%,~62\%]}\tabularnewline
    Bar  & One  & Curvy     & &\raL\bf\hfill88\%~{\tiny [66\%,~97\%]} &\raL\bf\hfill94\%~{\tiny [74\%,~99\%]} &\raL\bf\hfill100\%~{\tiny [86\%,~100\%]} &\raL\bf\hfill94\%~{\tiny [84\%,~98\%]}& &\raL   {\tiny$^{\vartriangle}_{\bullet}$}\hfill3.0$\pm$0.4 &\raL   {\tiny$^{\vartriangle}_{\bullet}$}\hfill3.2$\pm$0.5 &\raL   {\tiny$^{\vartriangle}_{\circ}$}\hfill3.1$\pm$0.5 &\raL   {\tiny$^{\vartriangle}_{\circ}$}\hfill3.1$\pm$0.3 &\raL    12\%~{\tiny [3\%,~34\%]}\tabularnewline
    Bar  & Alt. & Curvy     & &\raL   \hfill69\%~{\tiny [44\%,~87\%]} &\raL   \hfill75\%~{\tiny [51\%,~91\%]} &\raL   \hfill81\%~{\tiny [58\%,~94\%]} &\raL   \hfill75\%~{\tiny [62\%,~86\%]}& &\raL   {\tiny$^{\vartriangle}_{\bullet}$}\hfill2.5$\pm$0.5 &\raL   {\tiny$^{\vartriangle}_{\circ}$}\hfill2.6$\pm$0.5 &\raL   {\tiny$^{\vartriangle}_{\bullet}$}\hfill1.8$\pm$0.4 &\raL   {\tiny$^{\vartriangle}_{\circ}$}\hfill2.3$\pm$0.4 &\raL    6\%~{\tiny [1\%,~26\%]}\tabularnewline
    Leaf & One  & Straight  & &\raL   \hfill81\%~{\tiny [58\%,~94\%]} &\raL\bf\hfill94\%~{\tiny [74\%,~99\%]} &\raL   \hfill94\%~{\tiny [74\%,~99\%]} &\raL   \hfill90\%~{\tiny [79\%,~96\%]}& &\raL   {\tiny$^{\vartriangle}_{\circ}$}\hfill3.5$\pm$0.4 &\raL   {\tiny$^{\vartriangle}_{\bullet}$}\hfill3.6$\pm$0.5 &\raL   {\tiny$^{\vartriangle}_{\circ}$}\hfill3.2$\pm$0.5 &\raL\bf{\tiny$^{\vartriangle}_{\circ}$}\hfill3.5$\pm$0.3 &\raL    19\%~{\tiny [6\%,~42\%]}\tabularnewline
    Leaf & One  & Curvy     & &\raL\bf\hfill88\%~{\tiny [66\%,~97\%]} &\raL   \hfill88\%~{\tiny [66\%,~97\%]} &\raL\bf\hfill100\%~{\tiny [86\%,~100\%]} &\raL   \hfill92\%~{\tiny [81\%,~97\%]}& &\raL   {\tiny$^{\vartriangle}_{\bullet}$}\hfill3.5$\pm$0.5 &\raL   {\tiny$^{\vartriangle}_{\bullet}$}\hfill3.6$\pm$0.5 &\raL   {\tiny$^{\vartriangle}_{\circ}$}\hfill2.9$\pm$0.4 &\raL   {\tiny$^{\vartriangle}_{\circ}$}\hfill3.4$\pm$0.3 &\raL    19\%~{\tiny [6\%,~42\%]}\tabularnewline
    Leaf & Alt. & Curvy     & &\raL   \hfill69\%~{\tiny [44\%,~87\%]} &\raL   \hfill75\%~{\tiny [51\%,~91\%]} &\raL   \hfill94\%~{\tiny [74\%,~99\%]} &\raL   \hfill79\%~{\tiny [66\%,~89\%]}& &\raL\bf{\tiny$^{\vartriangle}_{\bullet}$}\hfill4.0$\pm$0.4 &\raL   {\tiny$^{\vartriangle}_{\circ}$}\hfill3.1$\pm$0.4 &\raL   {\tiny$^{\vartriangle}_{\bullet}$}\hfill2.0$\pm$0.5 &\raL   {\tiny$^{\vartriangle}_{\circ}$}\hfill3.0$\pm$0.3 &\raL    6\%~{\tiny [1\%,~26\%]}\tabularnewline
    \else
    Bar  & One  & Straight  & &\raL\bf\hfill88\% &\raL\bf\hfill94\% &\raL\bf\hfill100\% &\raL\bf\hfill94\%~{\tiny [84\%,~98\%]}& &\raL   {\tiny$^{\vartriangle}_{\bullet}$}\hfill2.4$\pm$0.3 &\raL\bf{\tiny$^{\vartriangle}_{\bullet}$}\hfill3.7$\pm$0.4 &\raL\bf{\tiny$^{\vartriangle}_{\bullet}$}\hfill4.4$\pm$0.2 &\raL\bf{\tiny$^{\vartriangle}_{\bullet}$}\hfill3.5$\pm$0.2 &\raL\bf 38\%~{\tiny [17\%,~62\%]}\tabularnewline
    Bar  & One  & Curvy     & &\raL\bf\hfill88\% &\raL\bf\hfill94\% &\raL\bf\hfill100\% &\raL\bf\hfill94\%~{\tiny [84\%,~98\%]}& &\raL   {\tiny$^{\vartriangle}_{\bullet}$}\hfill3.0$\pm$0.4 &\raL   {\tiny$^{\vartriangle}_{\bullet}$}\hfill3.2$\pm$0.5 &\raL   {\tiny$^{\vartriangle}_{\circ}$}\hfill3.1$\pm$0.5 &\raL   {\tiny$^{\vartriangle}_{\circ}$}\hfill3.1$\pm$0.3 &\raL    12\%~{\tiny [3\%,~34\%]}\tabularnewline
    Bar  & Alt. & Curvy     & &\raL   \hfill69\% &\raL   \hfill75\% &\raL   \hfill 81\% &\raL   \hfill75\%~{\tiny [62\%,~86\%]}& &\raL   {\tiny$^{\vartriangle}_{\bullet}$}\hfill2.5$\pm$0.5 &\raL   {\tiny$^{\vartriangle}_{\circ}$}\hfill2.6$\pm$0.5 &\raL   {\tiny$^{\vartriangle}_{\bullet}$}\hfill1.8$\pm$0.4 &\raL   {\tiny$^{\vartriangle}_{\circ}$}\hfill2.3$\pm$0.4 &\raL    6\%~{\tiny [1\%,~26\%]}\tabularnewline
    Leaf & One  & Straight  & &\raL   \hfill81\% &\raL\bf\hfill94\% &\raL   \hfill 94\% &\raL   \hfill90\%~{\tiny [79\%,~96\%]}& &\raL   {\tiny$^{\vartriangle}_{\circ}$}\hfill3.5$\pm$0.4 &\raL   {\tiny$^{\vartriangle}_{\bullet}$}\hfill3.6$\pm$0.5 &\raL   {\tiny$^{\vartriangle}_{\circ}$}\hfill3.2$\pm$0.5 &\raL\bf{\tiny$^{\vartriangle}_{\circ}$}\hfill3.5$\pm$0.3 &\raL    19\%~{\tiny [6\%,~42\%]}\tabularnewline
    Leaf & One  & Curvy     & &\raL\bf\hfill88\% &\raL   \hfill88\% &\raL\bf\hfill100\% &\raL   \hfill92\%~{\tiny [81\%,~97\%]}& &\raL   {\tiny$^{\vartriangle}_{\bullet}$}\hfill3.5$\pm$0.5 &\raL   {\tiny$^{\vartriangle}_{\bullet}$}\hfill3.6$\pm$0.5 &\raL   {\tiny$^{\vartriangle}_{\circ}$}\hfill2.9$\pm$0.4 &\raL   {\tiny$^{\vartriangle}_{\circ}$}\hfill3.4$\pm$0.3 &\raL    19\%~{\tiny [6\%,~42\%]}\tabularnewline
    Leaf & Alt. & Curvy     & &\raL   \hfill69\% &\raL   \hfill75\% &\raL   \hfill 94\% &\raL   \hfill79\%~{\tiny [66\%,~89\%]}& &\raL\bf{\tiny$^{\vartriangle}_{\bullet}$}\hfill4.0$\pm$0.4 &\raL   {\tiny$^{\vartriangle}_{\circ}$}\hfill3.1$\pm$0.4 &\raL   {\tiny$^{\vartriangle}_{\bullet}$}\hfill2.0$\pm$0.5 &\raL   {\tiny$^{\vartriangle}_{\circ}$}\hfill3.0$\pm$0.3 &\raL    6\%~{\tiny [1\%,~26\%]}\tabularnewline
    \fi
    \midrule
  \end{tabular}}
  {\parbox{\linewidth}{\scriptsize
\textit{Notes.}
      \textbf{Format:} mean [lower bound, upper bound].
      \textbf{Sides:} `Alt.' = Alternated sides.
      \textbf{Preference scores} range from 1: ``Strongly Disagree''; to 5: ``Strongly Agree.''
      \textbf{Values in bold print} are the best rates and scores, and $p$-values under 0.05.
      \textbf{Outliers} out of 5th and 95th percentiles: $\vartriangle$~= No outliers; $\blacktriangle{n}$~= $n$ outliers. 
      \textbf{Distribution:}\linebreak $\circ$~= Normal; $\bullet$~= Non-normal (by Shapiro--Wilk test).
      \textbf{Confidence intervals} are at a 95\% level, from Jeffreys Bayesian method for success rates, from a normal distribution for normally distributed scores, and from the percentile bootstrap method for non-normally distributed scores.
      \textbf{Acronyms:} `bos' = Bar One-sided Straight histogram, `boc' = Bar One-sided Curvy histogram, `bac' = Bar Alternated Curvy histogram, `los' = Leaf One-sided Straight histogram, `loc' = Leaf One-sided Curvy histogram, `lac' = Leaf Alternated Curvy histogram.
      \ifSTATSA
      \textbf{Statistical tests:}
      (0) None.
      (1) Cochran's Q test when six groups; McNemar's test when two groups.
      (2) Friedman's ANOVA when six groups; Wilcoxon signed-rank test when two groups.
      \textbf{Statistical significance marks} are: \dg{}~$p<.10$, *~$p<.05$, **~$p<.01$, ***~$p<.001$.
      \fi
    }
  }
\end{table*}

Assumption A1 on the more complicated reading of leaf decorations seems unverified, or only by a -3\% decrease for straight one-sided charts.
Globally, bar-like charts and leaf charts performed at equal average success rates (see \autoref{fig:results:online1:success}).
Therefore, decoding rates from leaf shapes must rely on efficient visual furled line analysis that preserves cognition capacities.

Assumption A2 on the slightly more complicated reading of curvy charts also appears unverified for bar-like charts (-0\% decrease) and leaf charts (+2\% increase) with the same one-sided anchorings.
Therefore, curvy trunks, as well as straight trunks, can be considered for design choices.

However, assumption A3 on the harder reading of alternated anchorings holds for both bar-like charts and leaf charts (with the same curvy trunks), even if the decreases are lower than expected (i.e., by only -20\% and -14\%, respectively).
The effect of not relying on visual persistence but cognition appears to affect readings.

Finally, assumption A4 on plant-like higher aesthetics appears to hold for leaf charts but not bar-like charts (see \autoref{fig:results:online1:aesthetics}).
Regarding leaf chars, the curvy alternated anchoring was rated with the highest aesthetics.

Encoding data series with leaf charts is possible because of high success rates, but only with one-sided anchorings, which performed as well as bar-like charts.
The alternated-leaf charts are the most aesthetic, but success rates are too low.
Hence, a new design iteration compelled compromising by an axial symmetry to the one-sided curvy-trunk version to get a two-sided anchoring that should look better like a plant.
Moreover, we assume that two-sided anchorings must be at least as readable as one-sided anchorings because they are merely an axial symmetry.
Consequently, the following section compares two alternative decorations for such curvy two-sided leaf charts.

\section{Second Online Study: Comparing Nature-Evocative Decorations}\label{sec:online2}

Even if participants favored the traditional bar histogram to perform the task during the first online study, the aim of this research remains a nature-evocative design.
Therefore, now that we have a better understanding of trunk, anchoring, and decoration factors, this section continues investigating only embellished charts by a second online study (N=25) involving only plant-like histograms with curvy trunks and two-sided (axis-symmetric) stems. 
The goal of this second online study is twofold: (1) to ensure the readability of curvy two-sided charts and (2) to confirm the final decoration choice that a physical prototype will subsequently implement.

This way, two nature-evocative designs are compared for curvy two-sided charts: one with growing bamboo sticks, called BambHISTO (see \autoref{fig:online2:histograms:bambhisto}), and one with unfurling plant leaves, called PlantHISTO (see \autoref{fig:online2:histograms:planthisto}). The labels inserted along trunks (at the anchoring points of BambHISTO's sticks and PlantHISTO's leaves) show the users that these sticks and leaves are encoding data. Moreover, even if some tiny leaves stand as small decorations on BambHISTO's sticks, only the sticks are of comparable size and number to PlantHISTO's leaves. Furthermore, only the sticks grow to make BambHISTO resemble a chart, not the tiny decorations, so that users will associate data encoding only with the sticks.

\def\factor{0.49}
\def\height{4cm}
\begin{figure}
  \begin{subfigure}[b]{0.49\linewidth}
    \centering
\includegraphics[clip,trim=2mm 0 2mm 0,clip,height=\height]{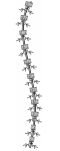}\includegraphics[clip,trim=2mm 0 2mm 0,clip,height=\height]{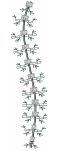}\includegraphics[height=\height]{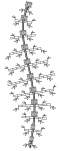}\includegraphics[clip,trim=0mm 0 0mm 0,clip,height=\height]{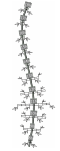}\includegraphics[clip,trim=0mm 0 0mm 0,clip,height=\height]{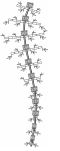}
    \caption{BambHISTO}
    \label{fig:online2:histograms:bambhisto}
    \Description{BambHISTO displays an eleven-hour energy variation used during the second online study.}
  \end{subfigure}
  \begin{subfigure}[b]{0.49\linewidth}
    \centering
\includegraphics[clip,trim=2mm 0 2mm 0,clip,height=\height]{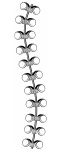}\includegraphics[clip,trim=2mm 0 2mm 0,clip,height=\height]{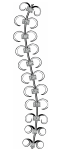}\includegraphics[height=\height]{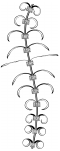}\includegraphics[clip,trim=0mm 0 0mm 0,clip,height=\height]{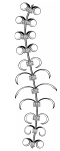}\includegraphics[clip,trim=0mm 0 0mm 0,clip,height=\height]{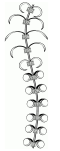}
    \caption{PlantHISTO}
    \label{fig:online2:histograms:planthisto}
    \Description{PlantHISTO displays an eleven-hour energy variation used during the second online study.}
  \end{subfigure}
\ifDRAFT
  \small{\it(More pictures were added during the revision)}
\fi
  \caption{
    Second Online Study -- The two low-fidelity histograms.
    Two dimensions are compared: decoration (bamboo vs leaf) and animation (growth vs unfurling).
  }
  \label{fig:online2:histograms}
\end{figure}
 
\subsection{Protocol}

Animations for six renewable-energy variations (see \autoref{tab:online2:variations}) were drawn and then layered into animated GIF files, which were integrated into online surveys\footnote{Lime Survey is an online survey web application (\href{https://www.limesurvey.org/}{https://www.limesurvey.org/}), which was chosen because enabling embbeding animated GIF files in questionnaires.}.
Increasing or decreasing stem shapes to show or hide energy variations was done by steps of $\pm$20~pt every 320~ms.
Those variations were used in four energy-shifting mini-scenarios where participants were asked when to use or recharge storage.
Twenty-five participants (lecturer-researchers and business owners: 12 women, 12 men, 0 non-binary, and 1 that prefers not to disclose, from 23 to 65 years old, 38.9 years old on average, SD=11.4) thus had to find the hours when to recharge (three mini-scenarios) and to discharge (one mini-scenario) laptop batteries at their office.
They were divided into two groups: one for each interaction condition.

\begin{table}
  \centering
  \caption{
    Second Online Study -- Design of six variations: three for each histogram.
    Those variations are inspired by real production data from solar panels registered at our lab.
  }~\label{tab:online2:variations}
  {\tablefontsize
  \begin{tabular}{lrrrrrclrrrrr}
    \cmidrule[0.5pt]{1-6}\cmidrule[0.5pt]{8-13}
    \it Histogram &\it Variation &\it Length &\it Start &\it Peak &\it End && \it Histogram &\it Variation &\it Length &\it Start &\it Peak &\it End \\
    \cmidrule{1-6}\cmidrule{8-13}
\ifHOURB
    BambHISTO  & \#1 & 9h &  9:00 & 13:00 & 17:59 &&    PlantHISTO & \#1 & 8h & 10:00 & 14:00 & 17:59\\
               & \#2 & 5h &  9:00 & 11:00 & 13:59 &&               & \#2 & 6h &  9:00 & 12:00 & 14:59\\  
               & \#3 & 5h & 13:00 & 16:00 & 17:59 &&               & \#3 & 4h & 14:00 & 16:00 & 17:59\\
\else
    BambHISTO  & \#1 & 9h &  9H & 13H & 17H       &&    PlantHISTO & \#1 & 8h & 10H & 14H & 17H\\
               & \#2 & 5h &  9H & 11H & 13H       &&               & \#2 & 6h &  9H & 12H & 14H\\  
               & \#3 & 5h & 13H & 16H & 17H       &&               & \#3 & 4h & 14H & 16H & 17H\\
\fi
    \cmidrule[0.5pt]{1-6}\cmidrule[0.5pt]{8-13}
  \end{tabular}
  }
\end{table}

\subsection{Assumptions}

We draw the following assumptions on readability, which directly affects performance (that is measured through success rates):

\begin{itemize}

\item[(A5)] \textbf{A two-sided anchoring should perform at least as good ($\pm$2~pts) as a one-sided anchoring.} According to the preceding results on bar-like charts of the first online study for one-sided anchorings, BambHISTOthe nature-evocative bar-like chart (BambHISTO) ought to perform as well as the preceding one-sided curvy bar chart (i.e., 94\%) and the unfurling-leaf chart (PlantHISTO) as the preceding one-sided curvy leaf chart (i.e., 92\%).
  
\item[(A6)] \textbf{Nature-evocative bar-like charts (for instance, BambHISTO) slightly outperform [+2\%, +4\%] unfurling-leaf charts (for instance, PlantHISTO).} According to the first online study's results on bar and bar-like charts with one-sided anchorings, reading straight bamboo sticks should slightly outperform the folded plant leaves.

\end{itemize}

\subsection{Results and Analysis}

\begin{figure}
  \def\factor{0.23}
  \begin{subfigure}[b]{\factor\linewidth} \centering
    \includegraphics[width=1.0\linewidth]{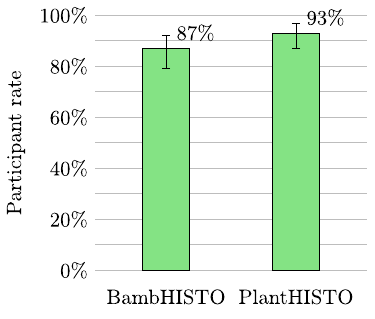}
    \caption{Task success.}
    \label{fig:results:online2:success}
    \Description{The bar chart shows the results of the task success of the second online study for BambHISTO and PlantHISTO.}
  \end{subfigure}
  \hspace{5mm}
  \def\factor{0.33}
  \begin{subfigure}[b]{\factor\linewidth} \centering
    \includegraphics[width=1.0\linewidth]{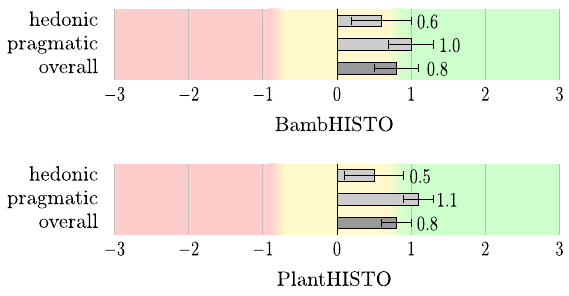}
    \caption{User Experience.}
    \label{fig:results:online2:ux}
    \Description{The bar chart shows the overall UEQ-S user experience scores of the second online study for BambHISTO and PlantHISTO.}
  \end{subfigure}
    {\parbox{0.80\linewidth}{\scriptsize
      \textit{Note.} \textbf{Error bars} are 95\% confidence intervals from a normal distribution for UEQ-S scores and from Jeffreys Bayesian method for success rates.
    }}
  \caption{Second Online Study -- Results (N=25).}
  \label{fig:results:online2}
\end{figure}

\autoref{tab:results:online2} shows the results.
PlantHISTO performed better than BambHISTO by a 6-point gap (see \autoref{fig:results:online2:success}).
\ifSTATSB
{on three of four tasks with statistical significance ($p$ < .003 by McNemar's test) and on overall (93\% overall success rate versus 87\%, respectively) even if a Wilcoxon signed-rank test suggests no statistical significance ($p$ = .111).}
\fi
Hedonic quality is similar in these two nature-evocating designs.
However, even if Bamboo sticks---which may recall traditional bars---were considered slightly more pragmatic, the overall UX qualities remain equally low (see \autoref{fig:results:online2:ux}), probably because of the low-fidelity level of the graphical animations.
Finally, despite minor differences, the plant design is retained because of better success rates.

\begin{table*}
  \caption{Second Online Study -- Results for the two plant-like histograms (N=25).}
  \label{tab:results:online2}
  {\tablefontsize
    \ifPERB
       \begin{tabular}{
        l
        p{44pt}p{44pt}p{40pt}p{40pt}p{40pt}
        p{0pt}
        p{35pt}p{35pt}p{35pt}
      }
    \else
      \begin{tabular}{
        l
        p{30pt}p{30pt}p{30pt}p{30pt}p{40pt}
        p{0pt}
        p{36pt}p{36pt}p{36pt}
      }
    \fi
    \midrule
    & \multicolumn{5}{l}{\tbh{Success Rates}} && \multicolumn{3}{l}{\tbh{UEQ-S Qualities}}\\
    \cmidrule{2-6}\cmidrule{8-10}
    \mcli{Histogram} & \mcli{T\ts{recharge1}} & \mcli{T\ts{recharge2}} & \mcli{T\ts{discharge1}} & \mcli{T\ts{recharge3}} & \mcli{Overall} && \mcli{Pragmatic} & \mcli{Hedonic} & \mcli{Overall} \\
    \midrule
    \ifPERB
    BambHISTO  &\raL    92\%~{\tiny [77\%,~98\%]} &\raL    92\%~{\tiny [77\%,~98\%]} &\raL    76\%~{\tiny [57\%,~89\%]} &\raL\bf 88\%~{\tiny [71\%,~96\%]} &\raL    87\%~{\tiny [79\%,~92\%]}& &\raL\bf{\tiny$^{\blacktriangle2}_{\circ}$}\hfill0.6$\pm$0.4 &\raL   {\tiny$^{\blacktriangle1}_{\circ}$}\hfill1.0$\pm$0.3 &\raL\bf{\tiny$^{\blacktriangle2}_{\circ}$}\hfill0.8$\pm$0.3\tabularnewline
    PlantHISTO  &\raL\bf 100\%~{\tiny [90\%,~100\%]} &\raL\bf 100\%~{\tiny [90\%,~100\%]} &\raL\bf 88\%~{\tiny [71\%,~96\%]} &\raL    84\%~{\tiny [66\%,~94\%]} &\raL\bf 93\%~{\tiny [87\%,~97\%]}& &\raL   {\tiny$^{\blacktriangle2}_{\circ}$}\hfill0.5$\pm$0.4 &\raL\bf{\tiny$^{\blacktriangle1}_{\circ}$}\hfill1.1$\pm$0.2 &\raL\bf{\tiny$^{\blacktriangle2}_{\circ}$}\hfill0.8$\pm$0.2\tabularnewline
    \else
    BambHISTO  &\raL    92\% &\raL    92\% &\raL    76\% &\raL\bf 88\% &\raL    87\%~{\tiny [79\%,~92\%]}& &\raL\bf{\tiny$^{\blacktriangle2}_{\circ}$}\hfill0.6$\pm$0.4 &\raL   {\tiny$^{\blacktriangle1}_{\circ}$}\hfill1.0$\pm$0.3 &\raL\bf{\tiny$^{\blacktriangle2}_{\circ}$}\hfill0.8$\pm$0.3\tabularnewline
    PlantHISTO  &\raL\bf 100\% &\raL\bf 100\% &\raL\bf 88\% &\raL    84\% &\raL\bf 93\%~{\tiny [87\%,~97\%]}& &\raL   {\tiny$^{\blacktriangle2}_{\circ}$}\hfill0.5$\pm$0.4 &\raL\bf{\tiny$^{\blacktriangle1}_{\circ}$}\hfill1.1$\pm$0.2 &\raL\bf{\tiny$^{\blacktriangle2}_{\circ}$}\hfill0.8$\pm$0.2\tabularnewline
    \fi
  \midrule
  \ifSTATSB
  & \multicolumn{5}{l}{\textbf{\textit{p}-values}} && \multicolumn{3}{l}{\textbf{\textit{p}-values}} \\
  \cmidrule{2-6}\cmidrule{8-10}
  &\ts{(1)} & \ts{(1)} & \ts{(1)} & \ts{(1)} & \ts{(2)} && \ts{(2)} & \ts{(2)} & \ts{(2)} \\
  \midrule
BH/PH & {\bf .000} ***   & {\bf .000} ***   & {\bf .002} **    & {\bf .000} ***   &      .111        &&      .369        &      .172        &      .375        \\
   \midrule
   \fi
  \end{tabular}}
  {\parbox{0.95\linewidth}{\scriptsize
      \textit{Notes.}
      \textbf{Format:} mean [lower bound, upper bound] for success rates,  mean$\pm$margin for scores.
      \textbf{UEQ-S scores} range from -3: ``Horribly bad''; to 3: ``Extremely good.''
      \textbf{Values in bold print} are the best rates and scores, and $p$-values under 0.05.
      \textbf{Outliers} out of 5th and 95th percentiles: $\vartriangle$~= No outliers; $\blacktriangle{n}$~= $n$ outliers. 
      \textbf{Distribution:} $\circ$~= Normal; $\bullet$~= Non-normal (by Shapiro--Wilk test).
      \textbf{Confidence intervals} are at a 95\% level, from Jeffreys Bayesian method for success rates, from a normal distribution for normally distributed scores, and from the percentile bootstrap method for non-normally distributed scores.
      \textbf{Acronyms:} `BH' = BambHISTO, `PH' = PlantHISTO.
      \ifSTATSB
      \textbf{Statistical tests:}
      (1) McNemar's when two groups.
      (2) Wilcoxon signed-rank test.
      \textbf{Statistical significance marks} are: \dg{}~$p<.10$, *~$p<.05$, **~$p<.01$, ***~$p<.001$.
      \fi
    }
  }
\end{table*}
 
Surprisingly, whereas the preceding bar-like charts with one-sided anchorings had high success rates during the first online study, the decorated bars of bamboo sticks had lower success rates during the second online study. After outperforming the leaf charts by an average of +3\%, bar-like charts now perform lower by -6\%.
Therefore, assumption A5 appears to hold for PlantHISTO (because of a +1~pt gap) but not for BambHISTO (because of a -7~pt gap).
Moreover, assumption A6 also does not appear to hold.
We hypothesize that this lower performance of BambHISTO comes from a less efficient visual reading of bamboo sticks' lengths because of the tiny decorations inserted all along.
Consequently, we assume in the remainder that assumption A5 remains valid even if not verified with BambHISTO.

Finally, after two online studies, exploring the design space leads to a single possible design of a plant-like chart to encode data series of renewable-energy rates: a curvy two-sided chart with unfurling leaves.
This chart cumulates both readability---for variations' endings and peak retrievals---as well as aesthetics---for more pleasant daily interaction.

\section{Physical Implementation: A Mechanism for Unfurling Leaves}\label{sec:implementation}

This section shares technical details about implementing the plant-like shape-changing interface PlantFORM (see \autoref{fig:teaser}), spanning from a folding mechanism to the holding structure and the mechatronics. Otherwise, the next section presents a user study comparing this prototype with three others.

\subsection{Folding Mechanism}

The folding mechanism is based on a pulling wire and a compression spring. During a first attempt, leaves were built from tiny wooden laser-cut parts that were mounted and articulated. Even if the mechanism worked, production and assembly times were relatively high---the new mechanism results from a second attempt restarting from an early paper-made medium-fidelity prototype.

\begin{figure}
  \begin{minipage}{4.2cm}\centering
  \begin{subfigure}[b]{1.0\linewidth}
\centering
    \includegraphics[width=4.1cm]{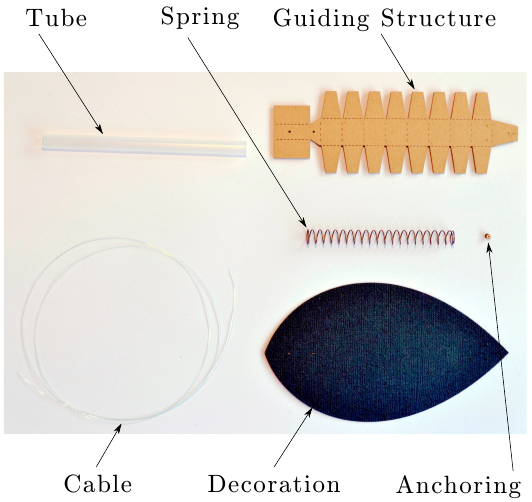}
    \caption{Spare parts of the mechanism.}
    \label{fig:leaf_mechanism:parts}
    \Description{The photograph shows the seven spare parts of the leaf mechanism: a tube, a spring, a guiding structure in cardboard, a Nylon cable, a decoration in paper shaped like a leaf, and a tiny anchoring cylinder.}
  \end{subfigure}
  \begin{subfigure}[b]{1.0\linewidth}
\centering
    \includegraphics[width=4.2cm]{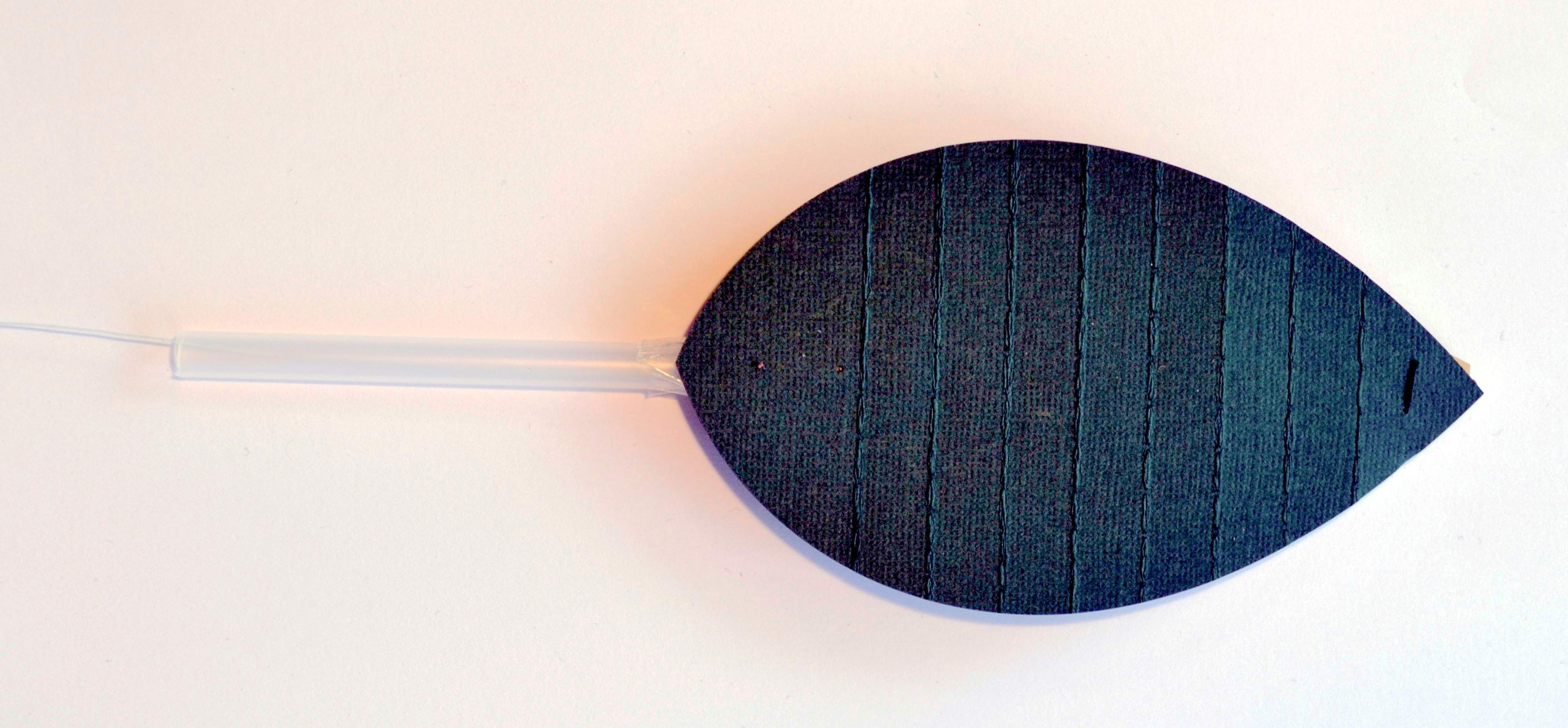}
    \includegraphics[width=4.2cm]{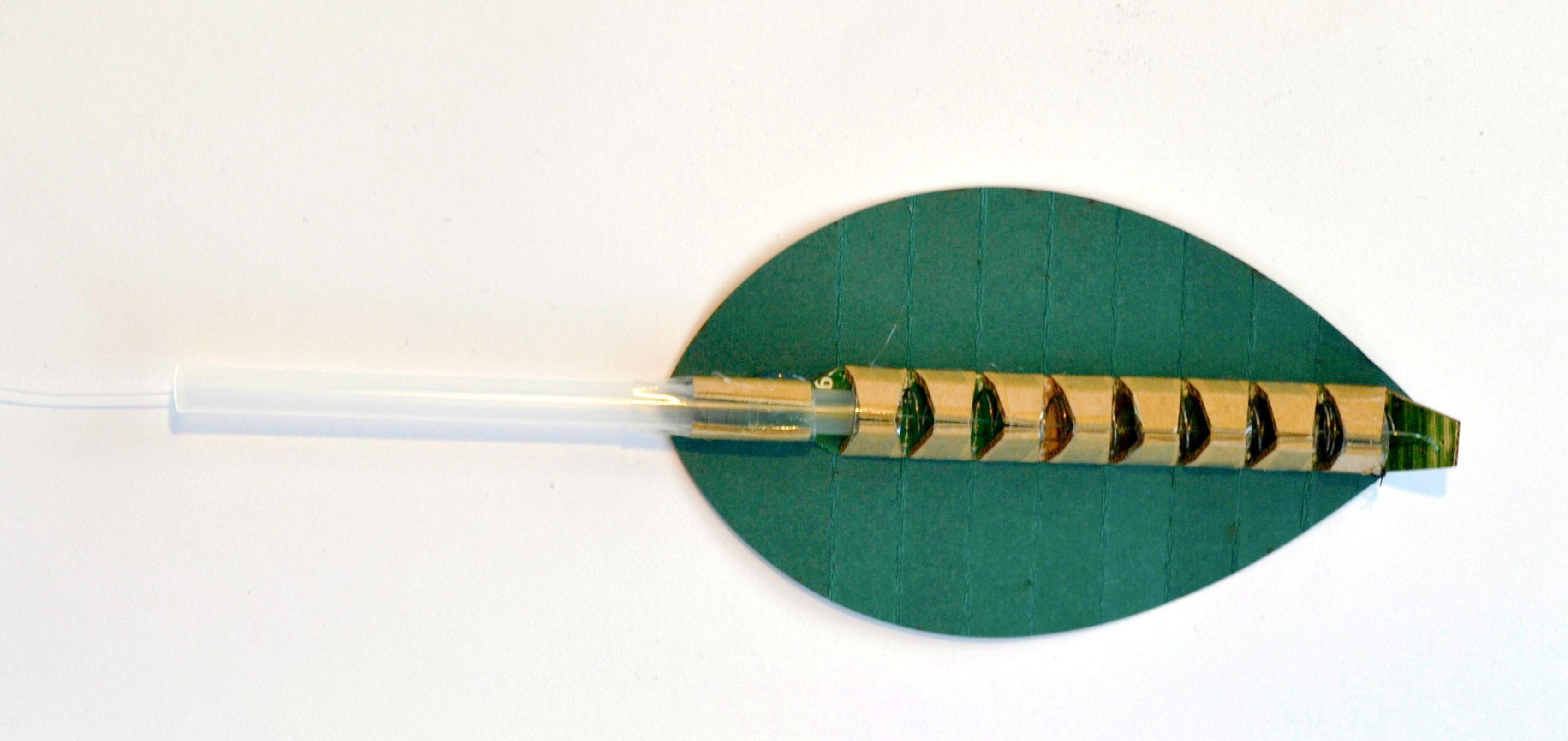}
    \caption{Assembly of a leaf.}
    \label{fig:leaf_mechanism:assembly}
    \Description{Two photographs show an assembled leaf from two viewpoints: from a top and a bottom view. The guiding structure is fixed to the decoration leaf; the tube is fixed at one end of the guiding structure, and the wire goes through the tube.}
  \end{subfigure}
  \end{minipage}\begin{minipage}{8.3cm}\centering
    \begin{subfigure}[b]{1.0\textwidth}
      \centering
      \includegraphics[width=8.04cm]{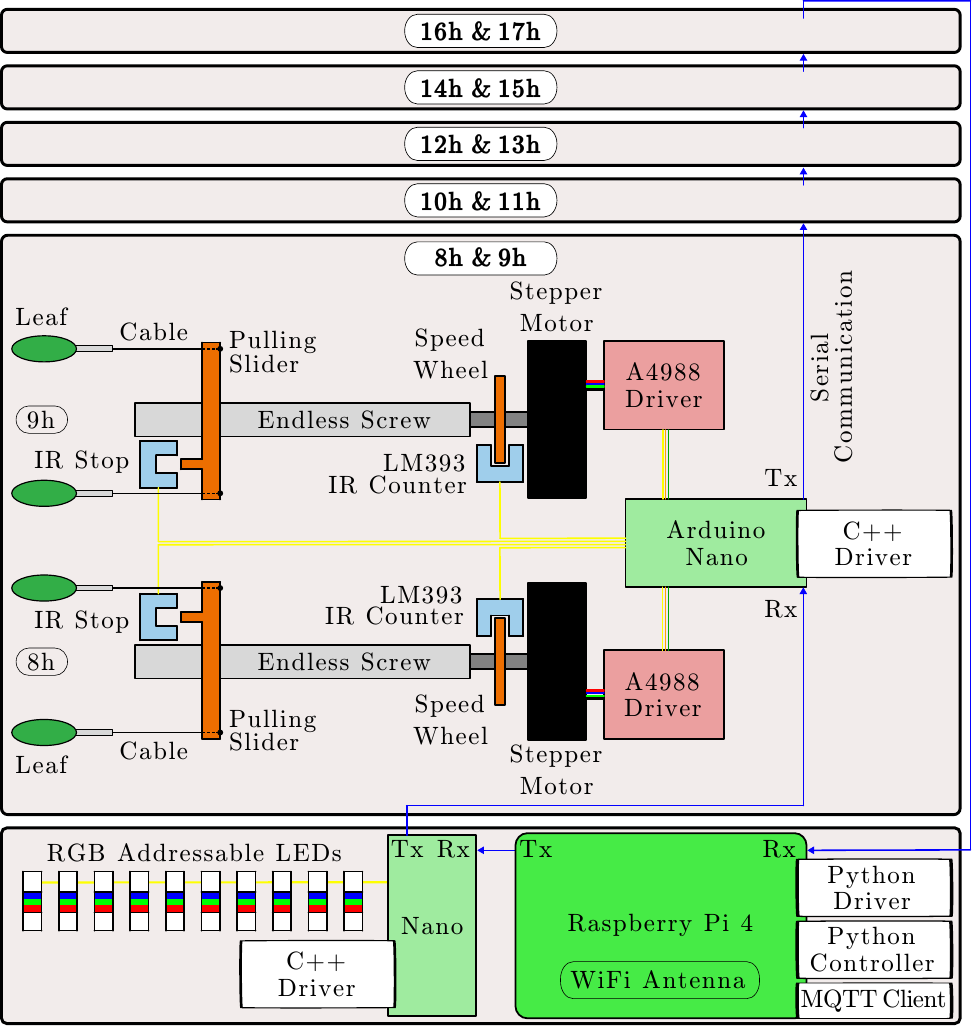}
      \caption{Diagram of the mechatronics and the pulling mechanism.}
      \label{fig:mechatronics}
    \end{subfigure}
  \end{minipage}\begin{minipage}{3.4cm}\centering
    \begin{subfigure}[b]{1.0\textwidth}
      \centering
      \includegraphics[clip,trim=2.5cm 0 3.6cm 0,clip,width=3.4cm]{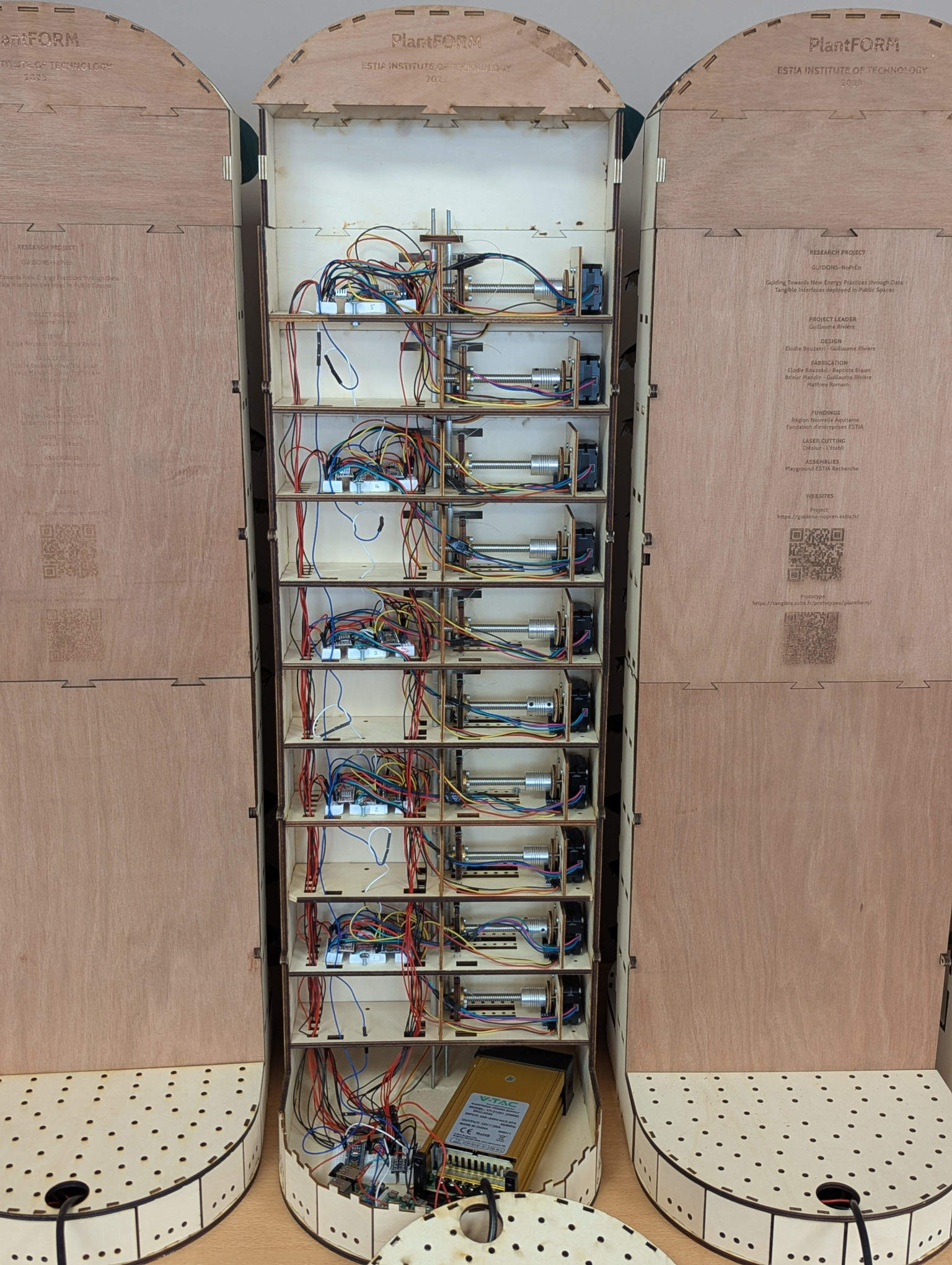} \caption{Rear wiring.}
      \label{fig:wiring}
    \end{subfigure}
  \end{minipage}
\ifDRAFT
  \small{\it(New figure added during the revision)}
\fi
  \caption{Implementing PlantFORM prototype. We provided a mechanical solution to leaf unfurling through a folding mechanism, based on a pulling cable, a spring, and a guiding structure. Pulling with endless screws enables keeping tension even when powering off the stepper motors. The wooden structure builds from laser cut parts that are assembled through flush-mounting. Prototyping the electronics relied on open-source hardware and software.}
  \label{fig:implementation}
  \Description{This diagram maps the electronics of PlantFORM and the pulling mechanism. A Raspberry Pi 4 single-board computer runs a Python Driver, a Python Controller, and a MQTT Client. This single-board computer is connected to a first Arduino Nano microcontroller board through serial communication. This microcontroller board is connected to a cluster of ten LEDs. This first microcontroller is connected through serial communication to a second one that controls the pulling mechanisms of the eight-hour and nine-hour leaves. Thus, the microcontroller is connected to two A4988 drivers that are connected to two stepper motors that are linked to two endless screws. Pulling sliders are fixed on the endless screws, which pull two cables from two leaves. Two LM393 IR sensors sense the stop position of the pulling slider and the steps of speed wheels that are fixed on stepper motors' shafts. The second microcontroller is connected through serial communication to the microcontroller of the ten-hour and eleven-hour leaves, and so on. Finally, the last microcontroller loops back serial communication to the single-board computer. All the micro-controller boards run a C++ Driver.}
\end{figure}

After several iterations---trying a jagged-cut plastic tube as a guiding and extension function---the final solution combines a laser-cut cardboard part as a guiding structure and a spring as an extension function (see \autoref{fig:leaf_mechanism:parts}). Leaves surfaces are laser cut from sheets of paper that were glued on the cardboard guides (but that can also be fixed with sewing threads). This final leaf mechanism is illustrated in \autoref{fig:leaf_mechanism:assembly}.

\subsection{Architecture and Mechatronics}

Finding the leaf mechanism required numerous iterations and lasted several months, whereas designing and implementing the structure and the pulling mechanism were straightforward and required only a few weeks.
The implementation involves laser-cut parts for the ten stages of the holding structure and off-the-shelf parts for mechanics (i.e., available commercially).
Parts were designed using CAD software (OnShape.com) and vector drawing software (Inkscape).

Actuation is done by ten stepper motors with 16 N.cm torque (StepperOnline 17HS08-1004S) controlled by ten drivers (A4988) and five microcontroller boards (Arduino Nano).
A sixth microcontroller drives ten digital RGB addressable LEDs.
One single-board computer (Raspberry Pi 4) controls the six Arduino boards through serial communication.
The single-board computer runs a program written in Python; the microcontrollers run a program written in Arduino C++.

Stepper motors are bound to endless screws that pull cables linked to the leaves: this solution enables keeping cable tension when turning off the motors---thus staying aligned with energy-saving goals.
The motor's position is controlled through LM393 infrared sensors: ten to notify stop positions and ten to count motor rotations.
On average, the motors' speed wheels count 216 rotation steps from the folded to the unfolded positions (this parameter was adjusted manually from 185 to 230, according to each leaf couple). This motion range is discretized in 11 positions, as illustrated in \autoref{fig:leaf_mechanism:unfolding}.

\def\indent{-1.90cm}
\def\width{1.95cm}
\def\vgap{4pt}
\begin{figure}
\hspace{\indent}\includegraphics[width=\width]{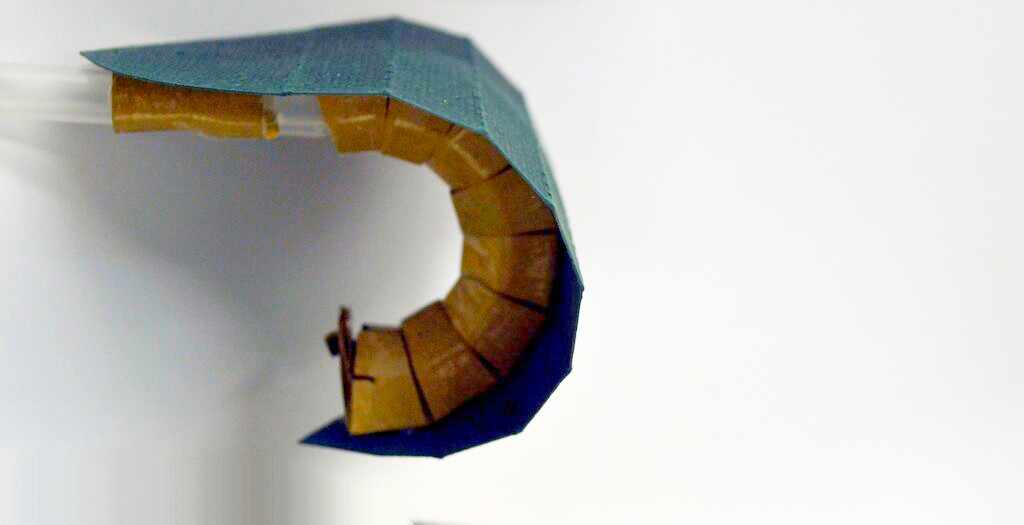}\hspace{-\width}0\hspace{\width}\includegraphics[width=\width]{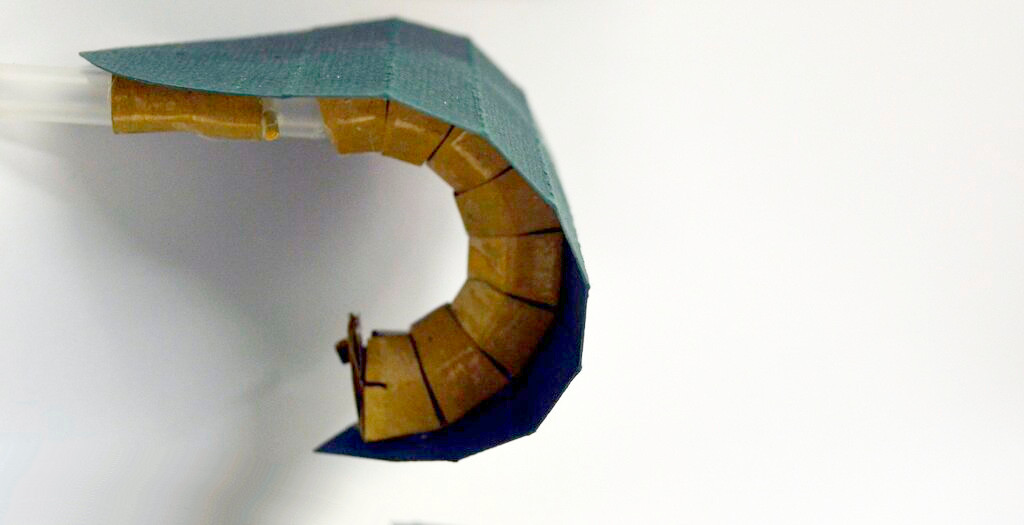}\hspace{-\width}1\hspace{\width}\includegraphics[width=\width]{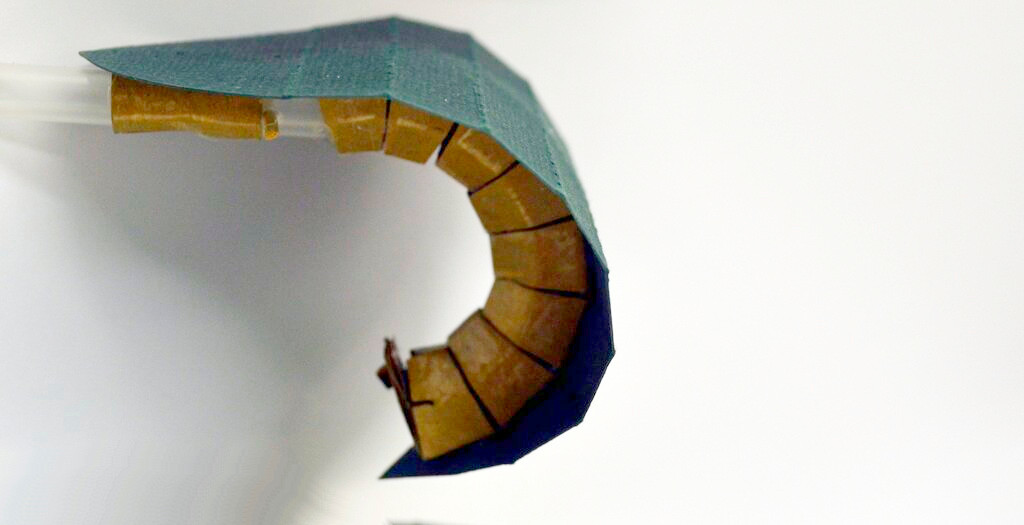}\hspace{-\width}2\hspace{\width}\includegraphics[width=\width]{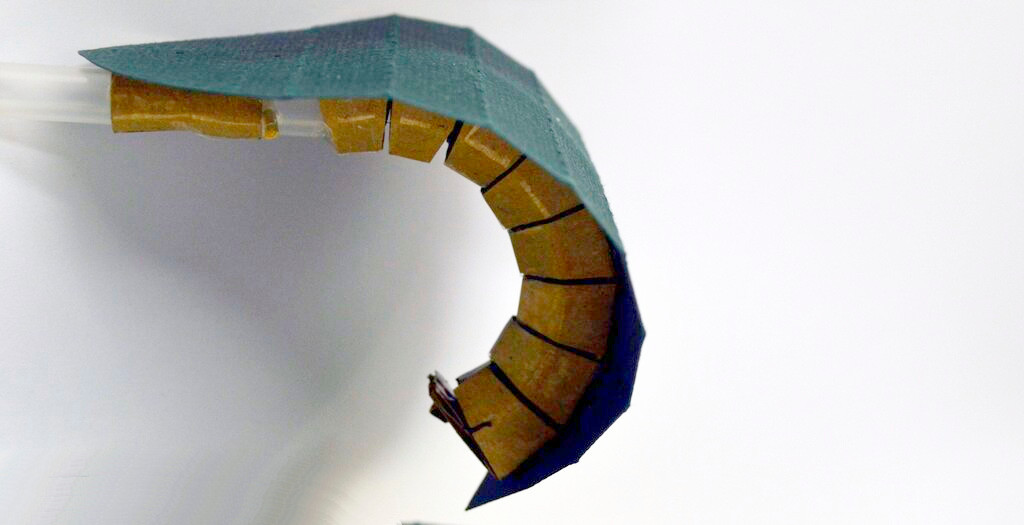}\hspace{-\width}3\hspace{\width}\includegraphics[width=\width]{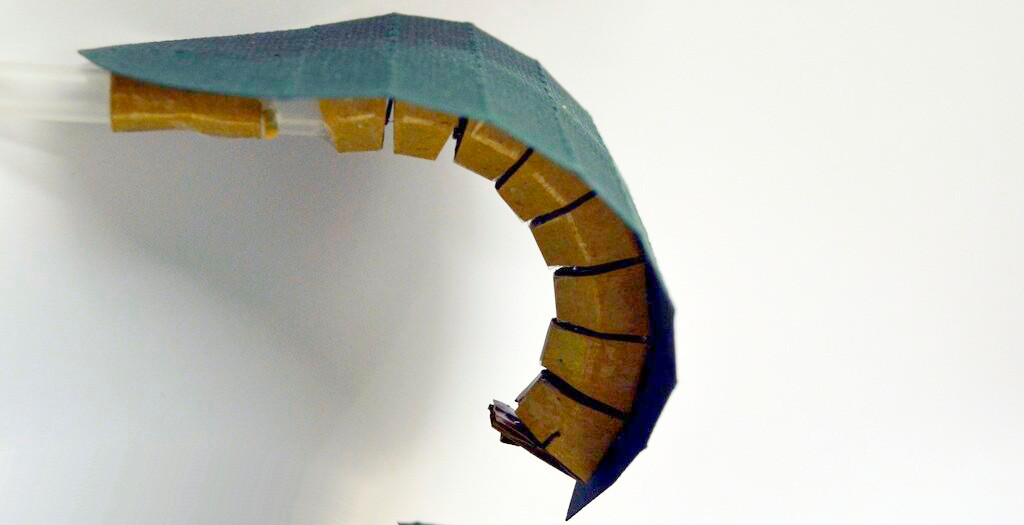}\hspace{-\width}4\hspace{\width}\includegraphics[width=\width]{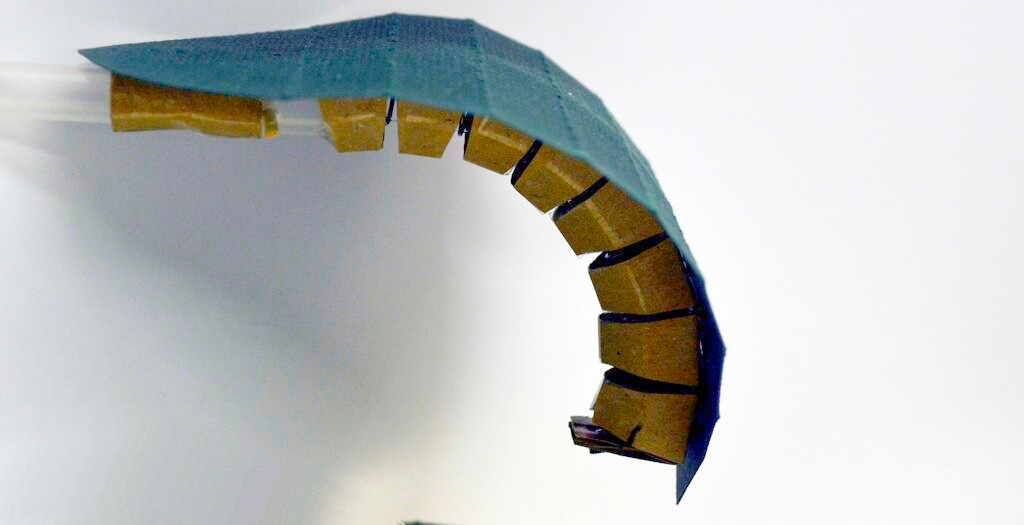}\hspace{-\width}5\hspace{\width}\\\vspace{\vgap}
\hspace{\indent}\includegraphics[width=\width]{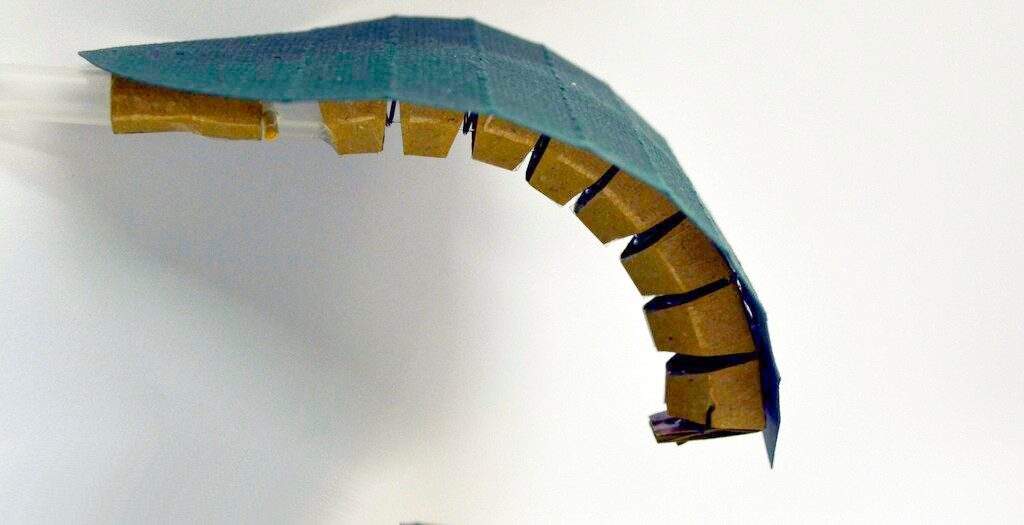}\hspace{-\width}6\hspace{\width}\includegraphics[width=\width]{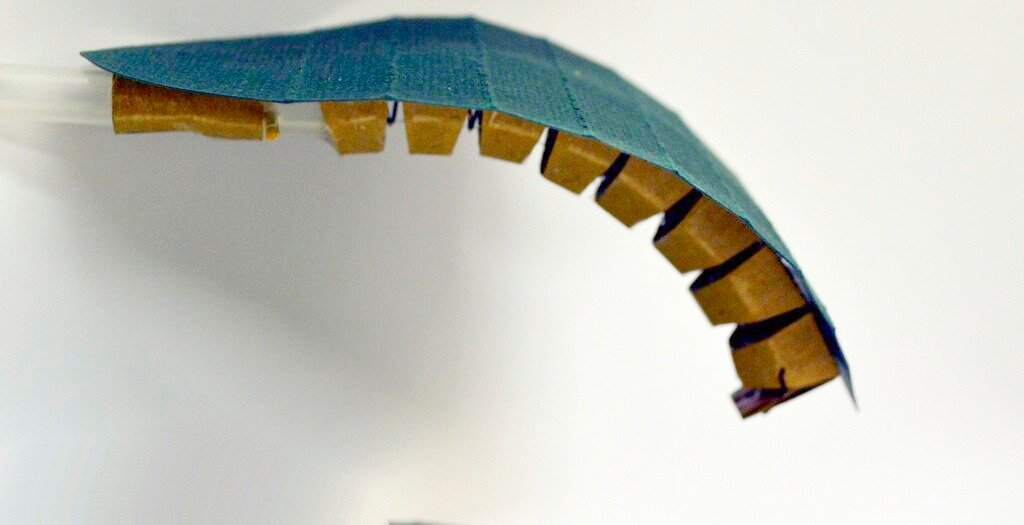}\hspace{-\width}7\hspace{\width}\includegraphics[width=\width]{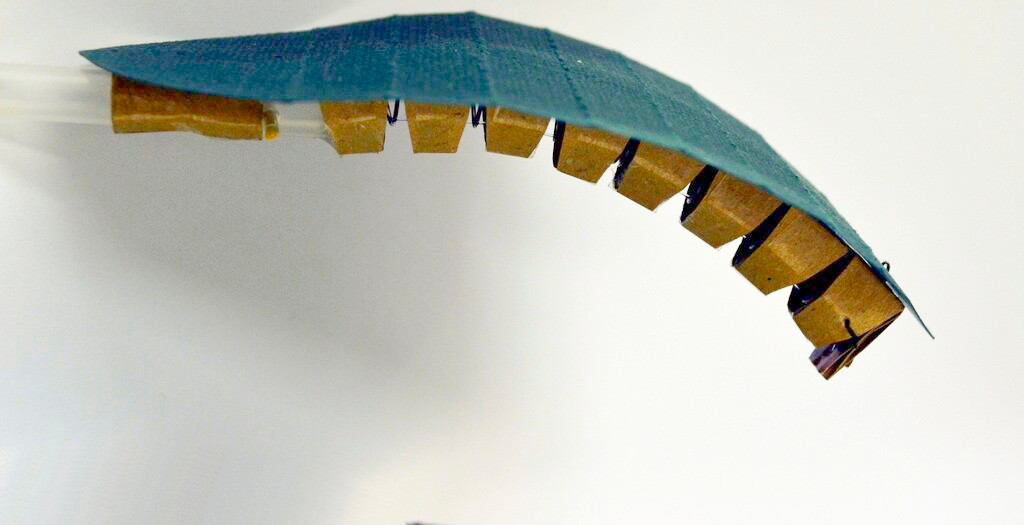}\hspace{-\width}8\hspace{\width}\includegraphics[width=\width]{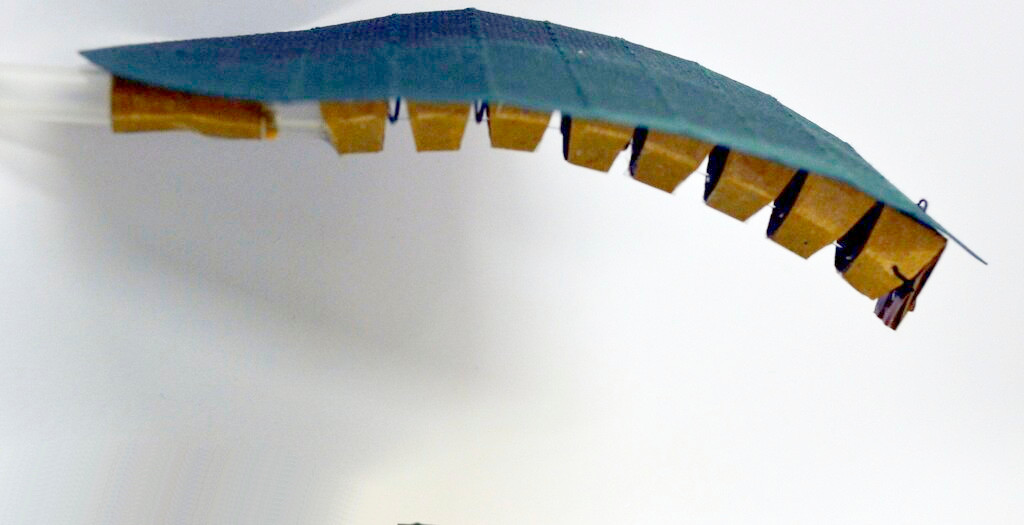}\hspace{-\width}9\hspace{\width}\includegraphics[width=\width]{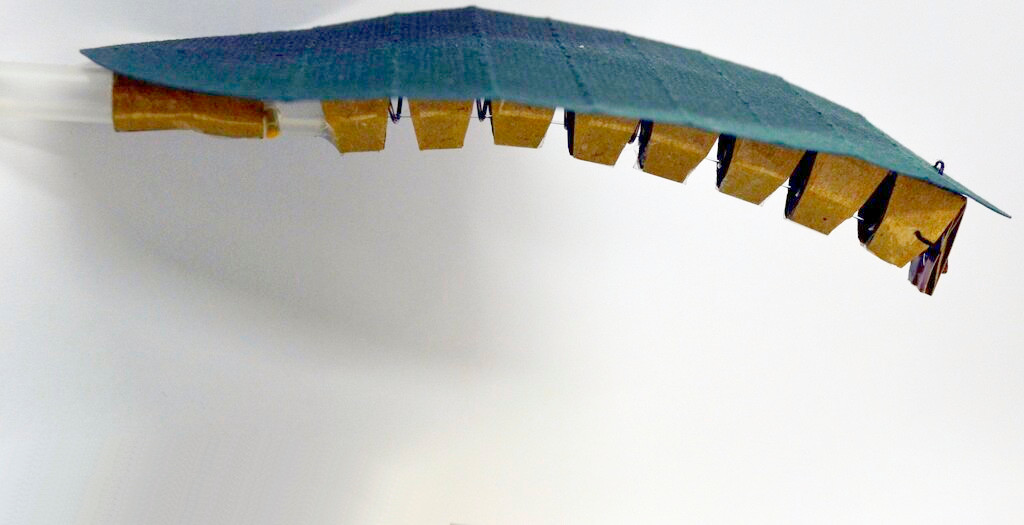}\hspace{-\width}10\hspace{\width}
\caption{The unfurling of a physical leaf prototype. Leaves are made of laser-cut cardboard and thick paper and are actuated using extension springs and pulling cables. Motion is discretized to eleven positions, from 0 to 10.}
    \label{fig:leaf_mechanism:unfolding}
    \Description{Eleven photographs show the folding of a physical leaf prototype, from position 0: ``folded'' to position 10: ``unfolded.''}
\end{figure}

Custom parts are wooden-made, produced by laser cutting and engraving, and assembled by flush-mounting. Electronics and mechanics parts are screwed.
The power supply is controlled by relays so that only the single-board computer and the LEDs are always on; microcontroller boards and stepper motors are power supplied only when actuating shapes.
The diagram of the mechatronics and the pulling mechanism is given in \autoref{fig:mechatronics}, and the structure and wiring are depicted in \autoref{fig:wiring}.

\def\footnotematerials{\footnote{Cardboard and paper are natural fiber materials that can be made resistant enough accordingly to usage, as in board games (e.g., jigsaw puzzles) and household decorations (e.g., lamp shades).}} 

Even if future implementations could use fewer electronics and mechanics, the development choices of this implementation already aspired for a low-tech attempt \cite{mirmalek2017lowtech}, as environmental impact is challenging for the future of tangible user interfaces \cite{holmquist2019future}. Therefore, the privileged choices were natural fibers for materials\footnotematerials{} and flush-mounting for assemblies (thus avoiding supplementary nuts and bolts). The following section presents a comparative user study involving this physical implementation of a plant-like chart.

\section{User Study: Comparing Modality and Material Attributes}\label{sec:userstudy} 

The two preceding online studies explored four dimensions of plant-like charts: trunk, anchoring, decoration, and animation.
This exploration led to a possible design optimizing readability and aesthetics.
The goal of this comparative user study (N=28) is to evaluate the implementation of this design and better understand the effects of two more dimensions when implementing embellished charts: modality (graphical versus physical) and materials (high-tech versus low-tech aspects), which may also impact aesthetics and user experience.

Thereby, readability, user experience, and user perception of naturalness and innovation are assessed on four prototypes of vertical charts, which are spread on a tech-trend continuum (see \autoref{fig:materials-continuum}), ranging from the low-tech aspects of a physical prototype using wood, cardboard, and paper to the high-tech aspects of another using illuminated translucent PMMA with visible technology, and passing by the half-combined aspects of two prototypes melting LCD screens and wood.

\def\factor{0.5}
\begin{figure}
\ifDRAFT
  \begin{minipage}{\factor\linewidth}
    \includegraphics[width=\linewidth]{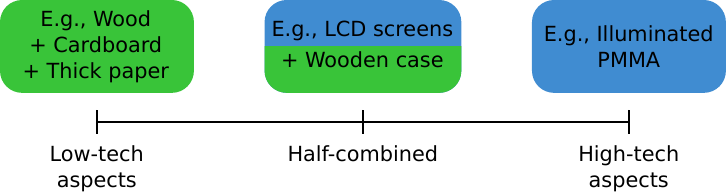}
  \end{minipage}\begin{minipage}{0.5cm}
  ~
  \end{minipage}\begin{minipage}{2.5cm}
    \small{\it(New figure added\linebreak during the revision)}
  \end{minipage}
\else
  \includegraphics[width=\factor\linewidth]{img/materials_continuum.pdf}
\fi
  \caption{
  Continuum of the perceived tech trends according to materials' choices.
  As packaging materials effects the perception of estimated environmental impacts \cite{koeniglewis2014consumers,lindh2016consumer,nguyen2020consumer}, the choice of materials is determinent in tech trend perception.
  }
  \label{fig:materials-continuum}
  \Description{TODO}
\end{figure}

\subsection{Interaction Conditions}

The physical leaf chart PlantFORM is compared with three other prototypes: a graphical leaf chart, a graphical bar-like chart, and a physical ring chart (see \autoref{fig:study:histograms}).

\def\factorA{0.19}
\def\factorB{0.19}
\def\factorC{0.19}
\def\factorD{0.19}
\def\height{5cm}
\begin{figure}
  \begin{subfigure}[b]{\factorA\linewidth}
    \centering
    \includegraphics[height=\height]{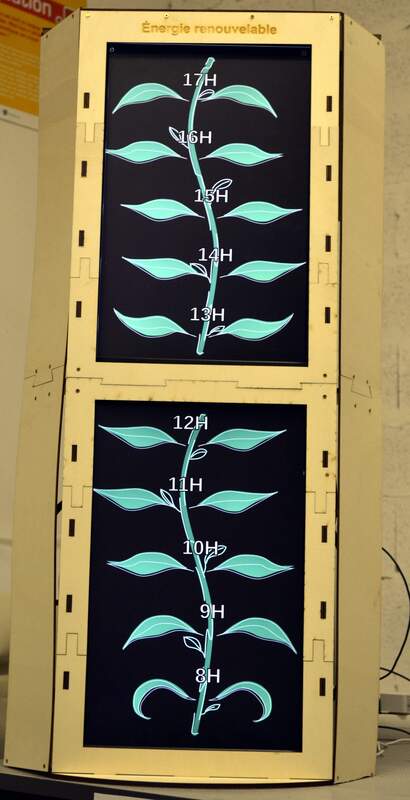}\\ {\scriptsize
    Graphical\\\vskip-3pt
    Wood/Screen/Pixels}
    \caption{PlantSCREEN}
    \label{fig:study:histograms:plantscreen}
    \Description{PlantSCREEN displays the first variation used during the user study.}
  \end{subfigure}\begin{subfigure}[b]{\factorB\linewidth}
    \centering
    \includegraphics[height=\height]{img/photos/plantform_var1.jpg}\\ {\scriptsize
    Physical\\\vskip-3pt
    Wood/Cardboard/Paper}
    \caption{PlantFORM}
    \label{fig:study:histograms:plantform}
    \Description{PlantFORM displays the first variation used during the user study.}
  \end{subfigure}\begin{subfigure}[b]{\factorC\linewidth}
    \centering
    \includegraphics[height=\height]{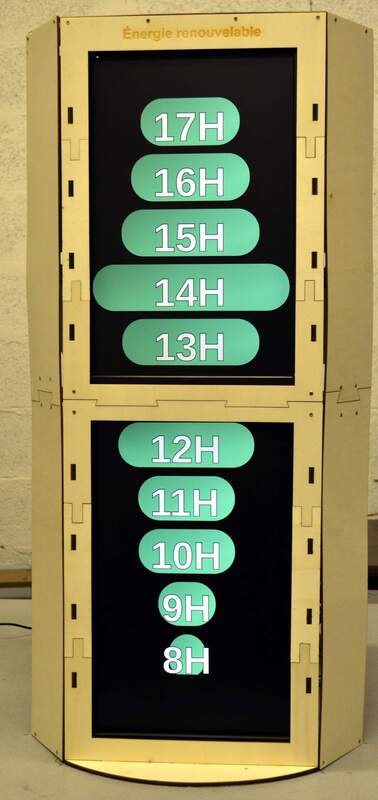}\\ {\scriptsize
    Graphical\\\vskip-3pt
    Wood/Screen/Pixels}
    \caption{CairnSCREEN}
    \label{fig:study:histograms:cairnscreen}
    \Description{CairnSCREEN displays the first variation used during the user study.}
  \end{subfigure}\begin{subfigure}[b]{\factorD\linewidth}
    \centering
    \includegraphics[height=\height]{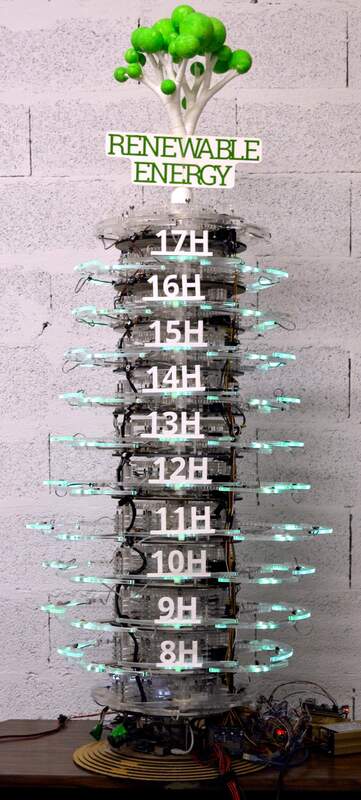}\\ {\scriptsize
    Physical\\\vskip-3pt
    Illuminated PMMA}
    \caption{CairnFORM}
    \label{fig:study:histograms:cairnform}
    \Description{CairnFORM displays the first variation used during the user study.}
  \end{subfigure}
  \caption{
    User Study -- The four high-fidelity histograms.
    Energy variations are encoded through shape-change without changing color gradient or intensity.
  }
  \label{fig:study:histograms}
\end{figure}
 
The other physical histogram is a modified version of CairnFORM \cite{daniel2019cairnform} that uses the same electronics and software as PlantFORM, thus reducing noise and vibrations thanks to smoother motion.
Ten expandable rings, made of translucent PMMA, are stacked and enlighted with green color using RGB LEDs through internal reflection.
Renewable energy rates are encoded through diameter expansion: the more expanded, the higher the rate.
From the user's viewpoint, rings' diameters grow as in classical bar charts.

The two graphical histograms are displayed as ambient graphical interfaces.
Whereas PlantSCREEN is a curvy two-sided unfurling-leaf histogram, CairnSCREEN is a straight axis-symmetric bar-like chart with smooth bar corners to mimic pebbles.
Pebbles' diameters grow as in traditional bar charts.
Hardware is a rebuilt version of CairnFORM's graphical version \cite{daniel2021cairnform}.
The histograms are displayed through two flat-screen LCD monitors of 19 inches and 1440 $\times$ 900 pixels, placed vertically, one under the other.
These two superimposed monitors are integrated into a casing---designed during two ideation sessions---to increase hardware aesthetics.
The final choice is a trapezoidal prism integrating the screens in their longer rectangular faces, which are fabricated with laser-cut wooden parts that are assembled by flush-mounting and pinning.

All four histograms were colored with unchanging green color, so only shape encodes energy rates.
Therefore, the ten RGB LEDs that are mounted along the trunk of PlantFORM were not used.
The motion of growing or unfurling shapes is discretized through 11 positions (i.e., the minimal position is coded 0, and the maximal position is coded 10).
When encoding rates for PlantFORM, CairnSCREEN, and CairnFORM, relative thresholding was computed (instead of linear mapping) according to the peak of the displayed variation (see \autoref{fig:data_encoding}).
This thresholding enables emphasizing energy variations' starts, peaks, and ends.
However, PlantSCREEN's leaf unfurling followed absolute thresholding (i.e., not relative to the peak rate).

\def\factor{0.7}
\begin{figure}
  \includegraphics[width=\factor\linewidth]{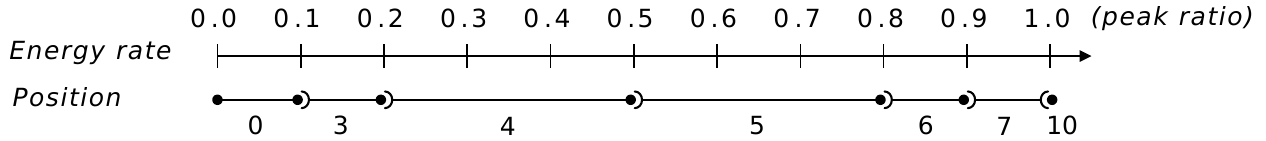}
  \caption{
    Data encoding relative to the peak.
    Energy rates are mapped to seven positions in a non-linear way that increases variation readability and emphasizes energy peaks.
}
  \label{fig:data_encoding}
  \Description{The relative rate encoding is illustrated by the correspondence between two parallel axes. The first axis is the energy rate that is expressed by a peak ratio from 0.0 to 1.0. The second axis is the position of the shape that encodes the rate according to peak ratio: position 0 in [0.0, 0.1], position 3 in ]0.1, 0.2], position 4 in ]0.2, 0.5], position 5 in ]0.5, 0.8], position 6 in ]0.8, 0.9], position 7 in ]0.9, 1.0], and position 10 when the peak ratio is 1.0.}
\end{figure}
 
For the needs of the user study, the prototypes were speeded up to sequentially display all ten hours from position 0 to position 10 in about 20$\pm$1 seconds.
Motion time was constant for the graphical histograms (i.e., the animation to display a rate always lasted 2 seconds, independently of the rate value), but relative to the rates for the physical histograms (i.e., the smaller the rate value, the shorter the animation time).
The speedup also increased the noise made by the shape-changing interfaces when animated.
The dimensions of the four histograms are compared in \autoref{tab:study:histograms_dimensions}.

\def\Position{Pos.}
\begin{table*}
  \caption{
    User Study -- Dimensions of the four histograms (in centimeters and inches).
    The dimensions are given for prototype and user study reproducibility concerns, as the size of the prototype influences readability and thus reading time and success.
}
  \label{tab:study:histograms_dimensions}
{\tablefontsize
  \begin{tabular}{
      l
      p{25pt}p{30pt}p{23pt}@{\hspace{1pt}}p{0pt}p{44pt}p{22pt}
      @{\hspace{4pt}}p{0pt}
      p{25pt}p{21pt}p{20pt}@{\hspace{1pt}}p{0pt}p{35pt}p{22pt}
    }
    \midrule

    & & \multicolumn{5}{l}{\textbf{\textit{Leaves/Bars/Rings}}}
    & && \multicolumn{5}{l}{\textbf{\textit{Leaves/Bars/Rings}}}\tabularnewline
    
    \cmidrule{3-7}\cmidrule{10-14}
    
    &   & \multicolumn{2}{l}{\textbf{\textit{Length/Diameter}}} && \multicolumn{2}{l}{\textbf{\textit{Thickness}}} && & \multicolumn{2}{l}{\textbf{\textit{Length/Diam.}}} && \multicolumn{2}{l}{\textbf{\textit{Thickness}}} \tabularnewline
    
    \cmidrule{3-4}\cmidrule{6-7}\cmidrule{10-11}\cmidrule{13-14}
    
    \textbf{\textit{Histogram}} & \textbf{\textit{Height}} &\it Position~0  &\it \Position{}~10 &&\it Position~0 &\it \Position{}~10 && \textbf{\textit{Height}} &\it Pos.~0  &\it \Position{}~10 &&\it Position~0 &\it \Position{}~10 \tabularnewline
    \midrule
    PlantSCREEN  &\raL 75.5~cm &\raL   4.7~cm &\raL 10.3~cm &&\raL  4.5~cm          &\raL 2.9~cm &&\raL 29.7"  &\raL   1.9" &\raL  4.1" &&\raL  1.8"          &\raL 1.1" \tabularnewline
    PlantFORM    &\raL 69.0~cm &\raL   6.5~cm &\raL 13.7~cm &&\raL  6.0$\pm$0.5~cm  &\raL 0.5~cm &&\raL 27.2"  &\raL   2.6" &\raL  5.4" &&\raL  2.4$\pm$0.2"  &\raL 0.2" \tabularnewline
    CairnSCREEN  &\raL 73.7~cm &\raL   0.0~cm &\raL 25.0~cm &&\raL  6.0~cm          &\raL 6.0~cm &&\raL 29.0"  &\raL   0.0" &\raL  9.8" &&\raL  2.4"          &\raL 2.4" \tabularnewline
    CairnFORM    &\raL 92.5~cm &\raL  35.0~cm &\raL 62.0~cm &&\raL  0.7~or~1.8~cm   &\raL 0.7~cm &&\raL 36.4"  &\raL  13.8" &\raL 24.4" &&\raL  0.3 or 0.7"   &\raL 0.1" \tabularnewline
    \midrule
  \end{tabular}
}{\parbox{\textwidth}{\centering
      \scriptsize
      \textit{Note.} Height is the physical height of the histograms (from the first hour to the last hour), not the entire structure.
  }}
\end{table*}

\subsection{Tasks}

Participants used the histograms to retrieve information for two energy storage tasks related to consumption shifts.
The first task \Trecharge{} required retrieving peak hours.
The second task \Tdischarge{} required retrieving energy variation's starting hours.
Tasks were executed using five energy variations (see \autoref{tab:study:variations}) adapted from past recorded production data (following the same curve shapes but with shifted peak hours).
Histograms displayed only one variation at a time; a displayed variation was closed before displaying the next one.
Sessions lasted between 45 minutes and one hour per participant.

\def\sepL{8pt}
\def\wLm{25pt} \def\wL{11pt} \def\wE{1pt}

\newcolumntype{A}{@{\hspace{2pt}}>{\raggedleft\arraybackslash}p{\wLm{}}}
\newcolumntype{B}{@{\hspace{2pt}}>{\raggedleft\arraybackslash}p{\wL{}}}
\newcolumntype{C}{@{\hspace{2pt}}>{\raggedleft\arraybackslash}p{18pt}}
\newcolumntype{D}{@{\hspace{2pt}}>{\raggedleft\arraybackslash}p{25pt}}

\begin{table*}
  \centering
  \caption{
    User Study -- Design of five variations on three days for the four interaction conditions.
    Whereas the first two days are day-long variations (8 or 10 hours long), the third day comprises three short variations (3, 4, or 5 hours long), thus soliciting the display capacities of prototypes. All those variations are inspired by real production data from solar panels registered at our lab.
  }~\label{tab:study:variations}
  {\tablefontsize
\ifHOURC
    \begin{tabular}{
        p{39pt}
@{\hspace{8pt}}p{4pt}@{\hspace{4pt}}p{24pt}@{\hspace{3pt}}p{16pt}@{\hspace{3pt}}p{17pt}@{\hspace{3pt}}p{22pt}
        @{\hspace{8pt}}p{4pt}@{\hspace{4pt}}p{22pt}@{\hspace{3pt}}p{17pt}@{\hspace{3pt}}p{17pt}@{\hspace{3pt}}p{22pt}
        @{\hspace{8pt}}p{4pt}@{\hspace{4pt}}p{13pt}@{\hspace{3pt}}p{16pt}@{\hspace{3pt}}p{17pt}@{\hspace{3pt}}p{17pt}
        @{\hspace{6pt}}p{1pt}@{\hspace{1pt}}p{13pt}@{\hspace{3pt}}p{17pt}@{\hspace{3pt}}p{17pt}@{\hspace{3pt}}p{17pt}
        @{\hspace{6pt}}p{1pt}@{\hspace{1pt}}p{13pt}@{\hspace{3pt}}p{17pt}@{\hspace{3pt}}p{17pt}@{\hspace{3pt}}p{17pt}
    } \else
    \begin{tabular}{
        p{39pt}
        @{\hspace{8pt}}p{4pt}@{\hspace{4pt}}p{24pt}@{\hspace{3pt}}p{16pt}@{\hspace{3pt}}p{17pt}@{\hspace{3pt}}p{22pt}
        @{\hspace{8pt}}p{4pt}@{\hspace{4pt}}p{22pt}@{\hspace{3pt}}p{17pt}@{\hspace{3pt}}p{17pt}@{\hspace{3pt}}p{22pt}
        @{\hspace{8pt}}p{4pt}@{\hspace{4pt}}p{9pt}@{\hspace{3pt}}p{11pt}@{\hspace{3pt}}p{13pt}@{\hspace{3pt}}p{13pt}
        @{\hspace{6pt}}p{1pt}@{\hspace{1pt}}p{9pt}@{\hspace{3pt}}p{11pt}@{\hspace{3pt}}p{13pt}@{\hspace{3pt}}p{13pt}
        @{\hspace{6pt}}p{1pt}@{\hspace{1pt}}p{9pt}@{\hspace{3pt}}p{11pt}@{\hspace{3pt}}p{13pt}@{\hspace{3pt}}p{13pt}
   }
\fi
    \midrule
    \multirow{2}{15mm}[-3pt]{\it Interaction\linebreak Condition} &&\multicolumn{4}{l}{\textbf{\textit{Monday}}} && \multicolumn{4}{l}{\textbf{\textit{Tuesday}}} && \multicolumn{14}{l}{\textbf{\textit{Wednesday}}} \\
    \cmidrule{3-6}\cmidrule{8-11}\cmidrule{13-26}
\ifHOURC
      &&\raR\it Len.\scriptsize{~(T)}&\raR\it Start &\raR\it Peak  &\raR\it End\scriptsize{~(t)}
      &&\raR\it Len.\scriptsize{~(T)} &\raR\it Start &\raR\it Peak  &\raR\it End\scriptsize{~(t)}
      &&\raR\it Len.&\raR\it Start &\raR\it Peak  &\raR\it End
      &&\raR\it Len.&\raR\it Start &\raR\it Peak  &\raR\it End
      &&\raR\it Len.&\raR\it Start &\raR\it Peak  &\raR\it End
      \tabularnewline
\else
      &&\raR\it Len.\scriptsize{~(T)}&\raR\it Start &\raR\it Peak  &\raR\it End\scriptsize{~(t)}
      &&\raR\it Len.\scriptsize{~(T)} &\raR\it Start &\raR\it Peak  &\raR\it End\scriptsize{~(t)}
      &&\raR\it L.&\raR\it S. &\raR\it P.  &\raR\it E.
      &&\raR\it L.&\raR\it S. &\raR\it P.  &\raR\it E.
      &&\raR\it L.&\raR\it S. &\raR\it P.  &\raR\it E.
      \tabularnewline
\fi
    \midrule
\ifHOURC
  \PS{} &&\raL 10h\scriptsize{~(20~s)} &\raL 8:00 &\raL 13:00 &\raL 17:59\scriptsize{~(8~s)} && 8h\scriptsize{~(16~s)} &\raL 10:00 &\raL 15:00 &\raL 17:59\scriptsize{~(4~s$^\ast$)} &&\raL 4h &\raL 8:00 &\raL 10:00 &\raL 11:59 &&\raL 3h &\raL 11:00 &\raL 12:00 &\raL 13:59 &&\raL 5h &\raL 13:00 &\raL 14:00 &\raL 17:59 \tabularnewline
  \PF{} &&\raL 10h\scriptsize{~(19~s)} &\raL 8:00 &\raL 12:00 &\raL 17:59\scriptsize{~(8~s)} && 8h\scriptsize{~(14~s)} &\raL 10:00 &\raL 16:00 &\raL 17:59\scriptsize{~(1.4~s)} &&\raL 5h &\raL 8:00 &\raL 9:00 &\raL 12:59 &&\raL 4h &\raL 12:00 &\raL 14:00 &\raL 15:59 &&\raL 3h &\raL 15:00 &\raL 16:00 &\raL 17:59 \tabularnewline
  \CS{} &&\raL 10h\scriptsize{~(20~s)} &\raL 8:00 &\raL 14:00 &\raL 17:59\scriptsize{~(6~s)} && 8h\scriptsize{~(16~s)} &\raL 10:00 &\raL 13:00 &\raL 17:59\scriptsize{~(8~s$^\ast$)} &&\raL 5h &\raL 8:00 &\raL 11:00 &\raL 12:59 &&\raL 3h &\raL 12:00 &\raL 13:00 &\raL 14:59 &&\raL 4h &\raL 14:00 &\raL 15:00 &\raL 17:59 \tabularnewline
  \CF{} &&\raL 10h\scriptsize{~(12~s)} &\raL 8:00 &\raL 11:00 &\raL 17:59\scriptsize{~(6~s)} && 8h\scriptsize{~(8~s)} &\raL 10:00 &\raL 14:00 &\raL 17:59\scriptsize{~(1.3~s)} &&\raL 3h &\raL 8:00 &\raL 9:00 &\raL 10:59 &&\raL 5h &\raL 10:00 &\raL 11:00 &\raL 14:59 &&\raL 4h &\raL 14:00 &\raL 16:00 &\raL 17:59 \tabularnewline
\else
  \PS{} &&\raL 10h\scriptsize{~(20~s)} &\raL 8H &\raL 13H &\raL 17H\scriptsize{~(8~s)} && 8h\scriptsize{~(16~s)} &\raL 10H &\raL 15H &\raL 17H\scriptsize{~(4~s$^\ast$)} &&\raL 4h &\raL 8H &\raL 10H &\raL 11H &&\raL 3h &\raL 11H &\raL 12H &\raL 13H &&\raL 5h &\raL 13H &\raL 14H &\raL 17H \tabularnewline
  \PF{} &&\raL 10h\scriptsize{~(19~s)} &\raL 8H &\raL 12H &\raL 17H\scriptsize{~(8~s)} && 8h\scriptsize{~(14~s)} &\raL 10H &\raL 16H &\raL 17H\scriptsize{~(1.4~s)} &&\raL 5h &\raL 8H &\raL 9H &\raL 12H &&\raL 4h &\raL 12H &\raL 14H &\raL 15H &&\raL 3h &\raL 15H &\raL 16H &\raL 17H \tabularnewline
  \CS{} &&\raL 10h\scriptsize{~(20~s)} &\raL 8H &\raL 14H &\raL 17H\scriptsize{~(6~s)} && 8h\scriptsize{~(16~s)} &\raL 10H &\raL 13H &\raL 17H\scriptsize{~(8~s$^\ast$)} &&\raL 5h &\raL 8H &\raL 11H &\raL 12H &&\raL 3h &\raL 12H &\raL 13H &\raL 14H &&\raL 4h &\raL 14H &\raL 15H &\raL 17H \tabularnewline
  \CF{} &&\raL 10h\scriptsize{~(12~s)} &\raL 8H &\raL 11H &\raL 17H\scriptsize{~(6~s)} && 8h\scriptsize{~(8~s)} &\raL 10H &\raL 14H &\raL 17H\scriptsize{~(1.3~s)} &&\raL 3h &\raL 8H &\raL 9H &\raL 10H &&\raL 5h &\raL 10H &\raL 11H &\raL 14H &&\raL 4h &\raL 14H &\raL 16H &\raL 17H \tabularnewline
\fi
  \midrule
  \end{tabular}
  }{\parbox{\textwidth}{\centering
      \scriptsize
      \textit{Notes.} `Len.' = Length. `T' = time to display the entire variation with the histogram. `t' = time from displaying the entire peak to displaying the last variation hour with the histogram. These variations' slopes are adapted from data measured at our laboratory. $^\ast$PlantSCREEN and CairnSCREEN reset all leaves and bars to position 0 before playing the second variation; PlantFORM and CairnFORM did not.
  }}
\end{table*}

\subsection{Procedure}

First, the user study leader presented the session's general course.
Then, participants were paid after filling out a consent form (including consent to be audio recorded).
Afterward, participants were asked to complete a questionnaire with demographic data.
Next, they were provided instructions on storage use and recharge to implement shifting strategies according to energy forecasts.
This presentation was followed by a training session involving \Trecharge{} and \Tdischarge{} using bar charts printed on sheets of paper.

Once the user study leader checked that the shifting strategy and the reading of energy variations (i.e., retrieving start, peak, and end points of slopes) were correctly understood, the participant was told that the audio recording was starting from that point onward.
Then, the leader presented each histogram one after the other.
However, no explanations were given on rate encoding, which is classical for the bar-like shapes but unusual for folding shapes (i.e., furled leaf = 0\% and unfurled leaf = 100\%).
Therefore, because the participants had to understand rate encoding by themselves, tasks also enabled observing the learnability of folded shapes.

Participants were told three mini-scenarios for each histogram, beginning at the start of a day when arriving at their office.
Participants had to answer the questions about five charging tasks (\Trecharge{}) and one discharging task (\Tdischarge{}) concerning days with single or multiple peaks.
They were asked to answer out loud as soon as they found the answer and then to write it down.
Once the use of a histogram ended, participants filled out the short ten-item questionnaire on the AttrakDiff scale \cite{hassenzahl_needs_2010}.

The study ended with a semi-structured interview through six questions.
The leader asked which histogram they think is the closest to nature, the most innovative, exciting, fascinating, and motivating.
Finally, they were asked which histogram they preferred to execute the tasks.

\subsection{Participants and Experimental Design}

The 28 participants (8 females and 20 males, from 20 to 50 years old, 32.5 years old on average, SD=8.6, and 5.4 years university degree on average, SD=2.3, ranging from 0 to 8 years university degree) were recruited by sending e-mails to university departments and campus companies, sticking posters, dropping off flyers, and asking people directly.

The user study took place over two weeks.
Two of the 30 participants initially recruited were removed due to incorrect completion of AttrakDiff questionnaires and misunderstanding of the exercises.
Participants' previous experience with physical shape-changing interfaces was rated from 1 `none' to 2 `little' (1.5 on average, SD=0.8, on a [1, 5] scale).
All of the participants reported working with laptops daily.
During this within-subjects user study, histograms' use was counterbalanced through four presentation orders (i.e., among four groups of users) by following a Latin square to prevent biases.

\subsection{Apparatus}

The experimental setup is depicted in \autoref{fig:apparatus}.
The four histograms were placed on two tables facing the participant's seat.
CairnFORM was further away than the other histograms in an attempt to preserve the same reading conditions as it is the taller and larger histogram.
The user study leader sat at the participant's left (see \autoref{fig:participant}).

\def\captionA{Top view of the apparatus (to scale). The setup is given to address the concerns of user study reproducibility and result interpretation. The prototypes' distance from the user was adjusted according to the size of leaves, bars, or rings and rotated to face the user.}
\def\descriptionA{The drawing shows a top view of the apparatus. The user sits at a table. The leader of the experiment sits to the left of the user. The four prototypes stand on two distant tables in front of the user. The distance between the user and the prototypes is shown, from left to right: PlantSCREEN stands at 2.2 m, PlantFORM at 2.4 m, CairnSCREEN at 1.9 m, and CairnFORM at 3.3 m.}
\def\captionB{One of the 28 participants, sat with the leader during the experiment. The distance and rotation of the prototypes to the user maintain similar readability among the four interaction conditions. Because CairnFORM is the largest of the four prototypes, it was placed the farthest.}
\def\descriptionB{The photograph shows the four prototypes in the background, displaying the last energy variations at the end of one participant's trials. The participant sits at the right of the foreground, and the experiment leader sits at the left. Some labels are stuck under the four prototypes, displaying their names.}
\begin{figure}
  \begin{subfigure}[b]{0.48\linewidth}
    \centering
    \includegraphics[width=0.6\linewidth]{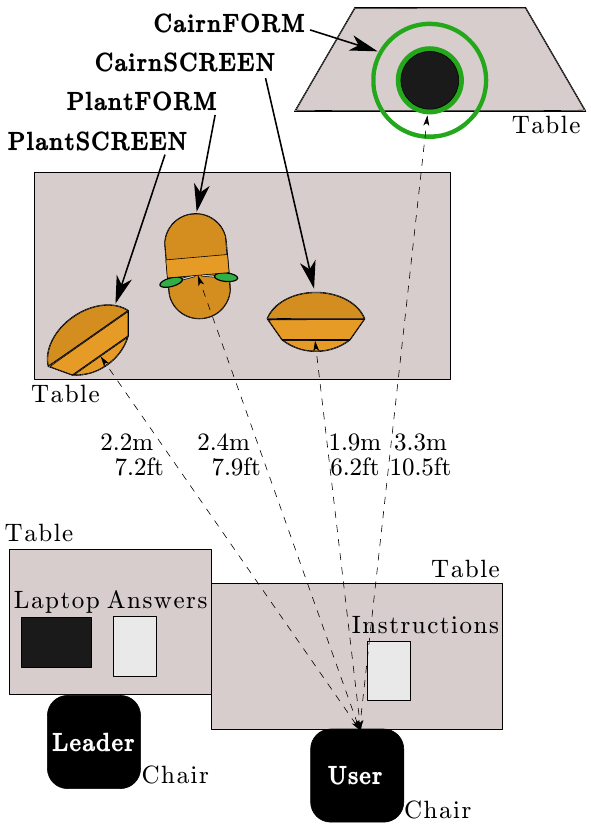} \caption{\captionA}
    \label{fig:apparatus}
    \Description{\descriptionA}
  \end{subfigure}\hspace{10pt}
  \begin{subfigure}[b]{0.48\linewidth}
    \centering
    \includegraphics[width=1.0\linewidth]{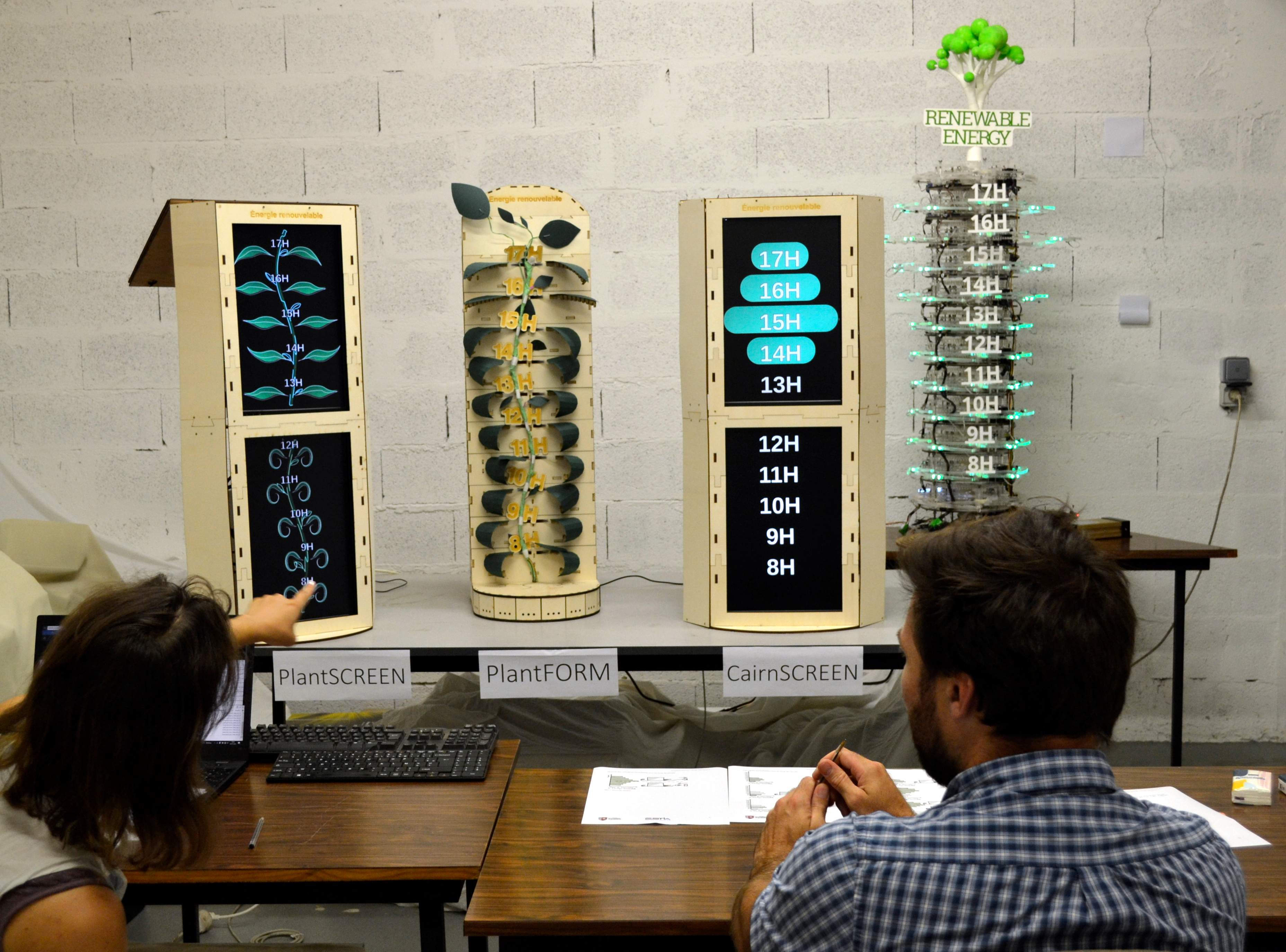}
    \caption{\captionB}
    \label{fig:participant}
    \Description{\descriptionB}
  \end{subfigure}
  \caption{User Study -- Experimental setup.}\label{fig:user_study}
\end{figure}
 
\subsection{Measures}

The measures included task success for efficacy and the ten-item AttrakDiff scale\footnote{The AttrakDiff construct was chosen to evaluate the high-fidelity prototypes because it emphasizes hedonic qualities---that are under interest---rather than pragmatic qualities \cite{laugwitz_construction_2008}.
The second online study resorted to the standardized and recently well-used UEQ construct \cite{diaz-oreiro_standardized_2019} because of low-fidelity black-and-white sketches; emphasizing hedonic qualities was odd.
The short versions of the questionnaires were chosen for duration reasons.} for user experience.
Tasks succeeded when participants answered correctly regarding storage use and recharge according to the displayed data series.
Also, efficacy included reading and interpreting charts correctly and answering correctly.

\subsection{Assumptions}

We draw four assumptions on the effect of modality and material factors on readability, which directly affects performance (that is measured through success rates) and user preferences:

\begin{itemize}

\item[(A7)] \textbf{Bar-like histograms perform slightly better [+2~pt, +5~pt] than plant-like histograms,} according to the results of the first online study, where the straight one-sided bar chart performed better than the curvy one-sided leaf chart.

\item[(A8)] \textbf{Modality does not affect chart readability.} Previous work has already observed very similar error rates between graphical and physical charts \cite{jansen_evaluating_2013}.
However, whereas previous work observed error rates close to zero \cite{jansen_evaluating_2013}, success rates for the present task, data, and histograms should approximate (by $\pm$5~pt) the results from preceding online studies.

\item[(A9)] \textbf{The choice of materials influences the tech trend perceived by users.} Beyond the plant-like design, using wood, cardboard, and thick paper must reinforce the user perception of PlantFORM being close to nature.
In some marketing studies, such an effect of materials is already observed in consumer perception of eco-friendly food packaging \cite{nguyen2020consumer}, where emotions prevail on rational evaluations \cite{koeniglewis2014consumers}.
  
For instance, consumers mostly gauge packaging's environmental impact by its materials \cite{lindh2016consumer} and understand well that paper-based packaging is more eco-friendly than plastic and metal \cite{lindh2016consumer}.
Moreover, consumer estimation of the environmental impact of paper-based packaging (including cardboard) aligns with scientific facts \cite{otto2021food}.
In the same way, influencing the tech trend using organic materials is acceptable if artifacts are in line with environmental sustainability.

\end{itemize}

\subsection{Results and Analysis}

This section reports the results of task performance, user experience, and user preferences for the four histograms.
\magenta{The results were analyzed through inferential statistics using non-parametric tests for dependent samples. The $p$-values and associated tests are reported in the results tables.} 

\subsubsection{Performance on Energy Storage Tasks}

\begin{figure*}
  \begin{subfigure}[b]{0.245\linewidth}
    \centering
    \includegraphics[width=1.0\linewidth]{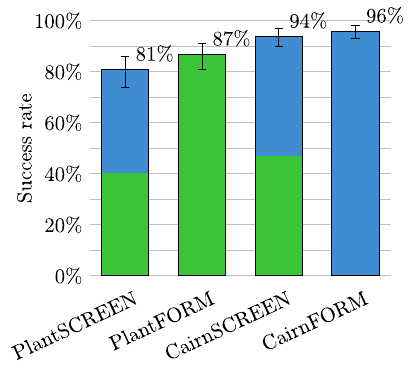}
    \caption{\tablefontsize Task overall success.}
    \label{fig:results:study:success}
    \Description{The bar chart shows means and 95\% CIs of task success of the user study for PlantSCREEN, PlantFORM, CairnSCREEN, and CairnFORM.}
  \end{subfigure}
\begin{subfigure}[b]{0.245\linewidth}
    \centering
    \includegraphics[width=1.0\linewidth]{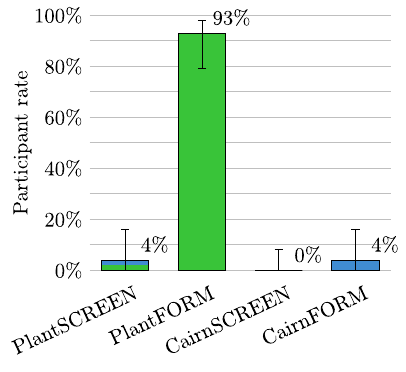}
    \caption{\tablefontsize Closest to nature.}
    \label{fig:results:study:nature}
    \Description{The bar chart shows means and 95\% CIs of participants' response rates for the closest to nature histogram at the end of the user study among PlantSCREEN, PlantFORM, CairnSCREEN, and CairnFORM.}
  \end{subfigure}
  \begin{subfigure}[b]{0.245\linewidth}
    \centering
    \includegraphics[width=1.0\linewidth]{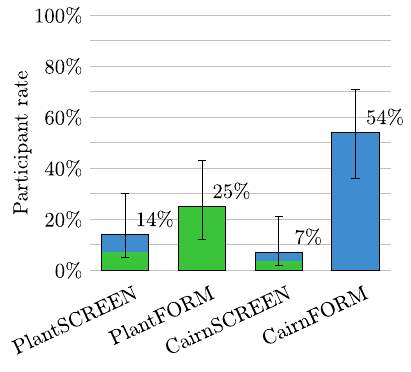}
    \caption{\tablefontsize The most innovative.}
    \label{fig:results:study:innovant}
    \Description{The bar chart shows means and 95\% CIs of participants' response rates for the most innovative histogram at the end of the user study among PlantSCREEN, PlantFORM, CairnSCREEN, and CairnFORM.}
  \end{subfigure}
  \begin{subfigure}[b]{0.245\linewidth}
    \centering
    \includegraphics[width=1.0\linewidth]{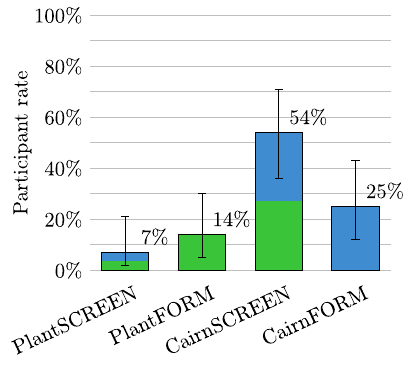}
    \caption{\tablefontsize The~most~task-adapted.}
    \label{fig:results:study:adapte}
    \Description{The bar chart shows means and 95\% CIs of participants' response rates for the most adapted histogram to tasks at the end of the user study among PlantSCREEN, PlantFORM, CairnSCREEN, and CairnFORM.}
  \end{subfigure}
    {\parbox{0.80\linewidth}{\centering
    \scriptsize
      \textit{Note.} \textbf{Error bars} are 95\% confidence intervals from a normal distribution for normally distributed scores, from the percentile bootstrap method for non-normally distributed scores, and from Jeffreys Bayesian method for rates.
    }}
  \caption{User Study -- Performance and some user preferences for the four histograms (N=28).}
  \label{fig:results:study}
\end{figure*}

\autoref{tab:results:study:success} reports the detailed success rates for the five recharge and one discharge trials, and \autoref{fig:results:study:success} shows the overall success rates for the six trials.
The overall success rates show an effect of the interaction conditions \magenta{($p$ = .002)} and that the two bar-like charts outperform the two plant-like charts with statistical significance \magenta{($p$ < .02 for all post hoc pairwise comparisons)}.

Overall, PlantFORM performed slightly better than PlantSCREEN but less than CairnSCREEN and CairnFORM.
The lower success rates of PlantSCREEN compared with PlantFORM are primarily due to unfurling beyond the horizontal line (e.g., see 13H on \autoref{fig:study:histograms:plantscreen}), as correlates with verbalization of some participants who pointed out confusion for the 100\% rate (i.e., excessive unfurling flaw).
Some participants also pointed out that differences between leaf positions must be increased, and one other that the zero position is difficult to discriminate (i.e., granularity flaw).

\begin{table*}
  \caption{User Study -- Task success (N=28).}
  \label{tab:results:study:success}
  {\tablefontsize
    \ifPERC
    \begin{tabular}{
        l
        @{}p{4pt}
        p{46pt}p{46pt}p{52pt}p{52pt}p{53pt}p{52pt}p{40pt}
      }
    \else
    \begin{tabular}{
        l
        @{}p{4pt}
        p{30pt}p{30pt}p{30pt}p{30pt}p{30pt}p{30pt}p{40pt}
      }
    \fi
    \midrule
     && \multicolumn{7}{l}{\tbh{Success Rates}}\\
    \cmidrule{3-9}
    \textit{Histogram} && \textit{T\ts{recharge1}}& \textit{T\ts{recharge2}}& \textit{T\ts{recharge3}}& \textit{T\ts{recharge4}}& \textit{T\ts{discharge1}}& \textit{T\ts{recharge5}}& \textit{Overall}\tabularnewline                        
    \midrule
    \ifPERC
    PlantSCREEN & &\raL    68\% {\tiny [50\%, 83\%]} &\raL    82\% {\tiny [65\%, 93\%]} &\raL\bf 100\% {\tiny [92\%, 100\%]} &\raL\bf 96\% {\tiny [84\%, 100\%]} &\raL    71\% {\tiny [53\%, 86\%]} &\raL    68\% {\tiny [50\%, 83\%]} &\raL    81\% {\tiny [74\%, 86\%]}\tabularnewline
    PlantFORM & &\raL    61\% {\tiny [42\%, 77\%]} &\raL    96\% {\tiny [84\%, 100\%]} &\raL    96\% {\tiny [84\%, 100\%]} &\raL    93\% {\tiny [79\%, 98\%]} &\raL    75\% {\tiny [57\%, 88\%]} &\raL\bf 100\% {\tiny [92\%, 100\%]} &\raL    87\% {\tiny [81\%, 91\%]}\tabularnewline
    CairnSCREEN & &\raL    93\% {\tiny [79\%, 98\%]} &\raL\bf 100\% {\tiny [92\%, 100\%]} &\raL\bf 100\% {\tiny [92\%, 100\%]} &\raL\bf 96\% {\tiny [84\%, 100\%]} &\raL\bf 89\% {\tiny [74\%, 97\%]} &\raL    86\% {\tiny [70\%, 95\%]} &\raL    94\% {\tiny [90\%, 97\%]}\tabularnewline
    CairnFORM & &\raL\bf 100\% {\tiny [92\%, 100\%]} &\raL\bf 100\% {\tiny [92\%, 100\%]} &\raL\bf 100\% {\tiny [92\%, 100\%]} &\raL\bf 96\% {\tiny [84\%, 100\%]} &\raL\bf 89\% {\tiny [74\%, 97\%]} &\raL    93\% {\tiny [79\%, 98\%]} &\raL\bf 96\% {\tiny [93\%, 98\%]}\tabularnewline
    \else
    PlantSCREEN & &\raL    68\% &\raL    82\% &\raL\bf 100\% &\raL\bf 96\% &\raL    71\% &\raL    68\% &\raL    81\%~{\tiny [74\%,~86\%]}\tabularnewline
    PlantFORM & &\raL    61\% &\raL    96\% &\raL    96\% &\raL    93\% &\raL    75\% &\raL\bf 100\% &\raL    87\%~{\tiny [81\%,~91\%]}\tabularnewline
    CairnSCREEN & &\raL    93\% &\raL\bf 100\% &\raL\bf 100\% &\raL\bf 96\% &\raL\bf 89\% &\raL    86\% &\raL    94\%~{\tiny [90\%,~97\%]}\tabularnewline
    CairnFORM & &\raL\bf 100\% &\raL\bf 100\% &\raL\bf 100\% &\raL\bf 96\% &\raL\bf 89\% &\raL    93\% &\raL\bf 96\%~{\tiny [93\%,~98\%]}\tabularnewline
    \fi
  \midrule
   && \multicolumn{7}{l}{\textbf{\textit{p}-values}} \\
  \cmidrule{3-9}
  && \ts{(1)} & \ts{(1)} & \ts{(1)} & \ts{(1)} & \ts{(1)} & \ts{(1)} & \ts{(2)} \\
  \midrule 
    PS/PF/CS/CF && {\bf .001} ***   & {\bf .010} *     &      .392        &      .875        &      .081  \dg{} & {\bf .000} ***   & {\bf .002} **    \\
    PS/PF &&      .117        & {\bf .000} ***   & {\bf .000} ***   & {\bf .000} ***   & {\bf .016} *     & {\bf .002} **    &      .135        \\
    PS/CS && {\bf .004} **    & {\bf .000} ***   & {\bf .000} ***   & {\bf .000} ***   & {\bf .003} **    & {\bf .009} **    & {\bf .012} *     \\
    PS/CF && {\bf .002} **    & {\bf .000} ***   & {\bf .000} ***   & {\bf .000} ***   & {\bf .003} **    & {\bf .004} **    & {\bf .001} **    \\
    PF/CS && {\bf .014} *     & {\bf .000} ***   & {\bf .000} ***   & {\bf .000} ***   & {\bf .001} **    & {\bf .000} ***   & {\bf .013} *     \\
    PF/CF && {\bf .006} **    & {\bf .000} ***   & {\bf .000} ***   & {\bf .000} ***   & {\bf .001} **    & {\bf .000} ***   & {\bf .004} **    \\
    CS/CF && {\bf .000} ***   & {\bf .000} ***   & {\bf .000} ***   & {\bf .000} ***   & {\bf .000} ***   & {\bf .000} ***   &      .187        \\
  \midrule
  \end{tabular}}
  {\parbox{0.85\linewidth}{\scriptsize
      \textit{Notes.}
      \textbf{Format:} mean [lower bound, upper bound].
      \textbf{Values in bold print} are the best rates and $p$-values under 0.05.
      \textbf{Confidence intervals} are at a 95\% level from Jeffreys Bayesian method for success rates.
      \textbf{Acronyms:} `PS' = PlantSCREEN, `PF' = PlantFORM, `CS' = CairnSCREEN, `CF' = CairnFORM.
      \textbf{Statistical tests:}
      (1) Cochran's Q test when four groups; McNemar's test when two groups.
      (2) Friedman's ANOVA when four groups; Wilcoxon signed-rank test when two groups.
      \textbf{Statistical significance marks} are: \dg{}~$p<.10$, *~$p<.05$, **~$p<.01$, ***~$p<.001$.
    }
  }
\end{table*}
 
The initial assumption on two-sided anchoring performance, which was assumed from one-sided anchoring, is again verified for bar-like charts.
Indeed, the success rates of CairnSCREEN and CairnFORM remain similar to the one-sided straight bar chart of the first online study, by a 1-point gap.
However, the lower success rates of PlantSCREEN and PlantFORM, compared with the one-sided curvy plant-like chart of the first online study, fail to repeat this initial observation because of higher 9- and 4-point gaps, even if the second online study did with PlantHISTO (by a 1-point gap).
Nevertheless, this failure can be explained by implementation flaws (i.e., excessive unfurling of PlantSCREEN and granularity of leaf encoding), which must have degraded the performance of leaf charts.
Therefore, working with the initial assumption remains valid also for plant-like charts.

Assumption A7 on data encoding's effect on reading performance appears to hold, even if the gap is more significant than expected. This gap increase is due to the downgraded performance of plant-like charts caused by implementation flaws. Improving the implementation of plant-like charts should bring higher success rates, which would, however, stay slightly below bar-like charts.

Assumption A8 on the absence of modality effect on reading performance also appears to hold, even if the gap is not precisely 0~points.
Indeed, experimental variations can explain the 2-point gap for bar-like charts.
Moreover, the implementation flaws can explain an inflated 6-point gap for plant-like charts.
Therefore, the effect of modality on success rates must remain negligible.

\subsubsection{Learnability of Leaf Encoding}

\begin{figure*}
  \begin{subfigure}[b]{0.245\linewidth}
    \centering
    \includegraphics[width=1.0\linewidth]{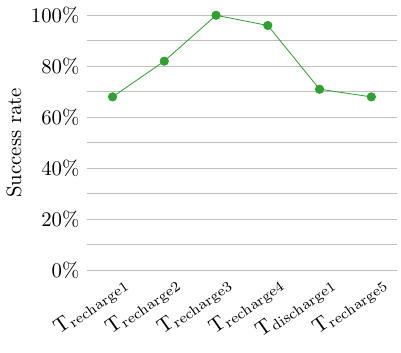}
    \caption{\tablefontsize PlantSCREEN.}
    \label{fig:learnability:study:success:ps}
    \Description{A multiline depicts the tendency of the success rates for each of the six recharging and discharging tasks of the user study for the PlantSCREEN prototype. The line grows from 68\% to over 96\% but then declines to less than 71\%.}
  \end{subfigure}
  \begin{subfigure}[b]{0.245\linewidth}
    \centering
    \includegraphics[width=1.0\linewidth]{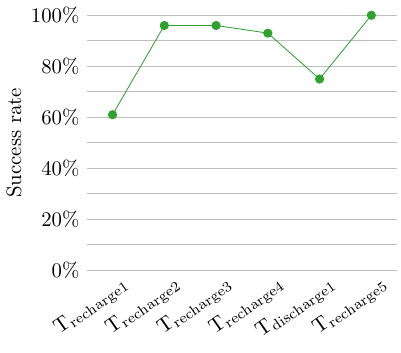}
    \caption{\tablefontsize PlantFORM.}
    \label{fig:learnability:study:success:pf}
    \Description{A multiline depicts the tendency of the success rates for each of the six recharging and discharging tasks of the user study for the PlantFORM prototype. The line grows very quickly from 61\% to a plateau beyond 93\% then declines to 75\% (i.e., the discharging task), and finally ends at 100\%.}
  \end{subfigure}
  \begin{subfigure}[b]{0.245\linewidth}
    \centering
    \includegraphics[width=1.0\linewidth]{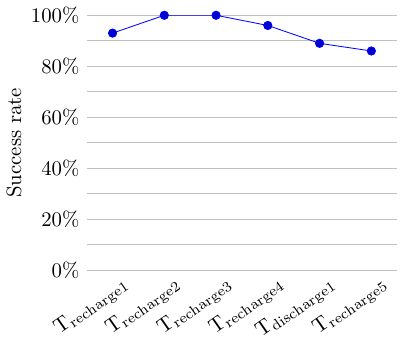}
    \caption{\tablefontsize CairnSCREEN.}
    \label{fig:learnability:study:success:cs}
    \Description{A multiline depicts the tendency of the success rates for each of the six recharging and discharging tasks of the user study for the CairnSCREEN prototype. The line is relatively flat and high overall (close to 100\%), but decreases after half of the tasks.}
  \end{subfigure}
  \begin{subfigure}[b]{0.245\linewidth}
    \centering
    \includegraphics[width=1.0\linewidth]{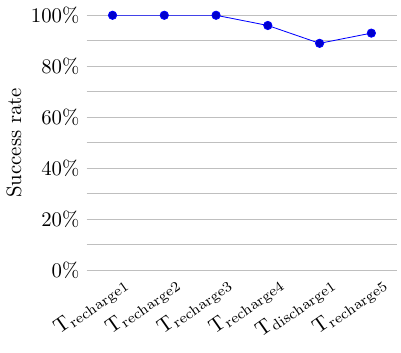}
    \caption{\tablefontsize CairnFORM.}
    \label{fig:learnability:study:success:cf}
    \Description{A multiline depicts the tendency of the success rates for each of the six recharging and discharging tasks of the user study for the CairnFORM prototype. The line is relatively flat and very high overall (close to 100\%) but decreases slightly after half of the tasks.}
  \end{subfigure}
  \caption{User Study -- Success rates per task for the four histograms (N=28).}
  \label{fig:learnability:study}
\end{figure*}

Understanding rate encoding by folding shapes required a short learning phase (see $T_{recharge1}$, $T_{recharge2}$, and $T_{recharge3}$ in \autoref{fig:learnability:study}).
As no explanations were given on the unusual encoding of rates with folding shapes, a learning effect must explain the low start of PlantSCREEN and PlantFORM on $T_{recharge1}$ and then increase on $T_{recharge2}$ and $T_{recharge3}$, compared with the growing shapes of CairnSCREEN and CairnFORM that immediately reached high rates from $T_{recharge1}$. Learning of folded shapes encoding seems, however, to happen very quickly.

\subsubsection{User Experience and Preferences}

The results of the four AttrakDiff qualities and the overall scores are reported in \autoref{tab:results:study:ux}, and the scores for the ten items of the scale are illustrated in \autoref{fig:results:study:ux_detail}.
The user experience is quite similar over the four histograms, with medium overall AttrakDiff scores and without an effect of the interaction conditions \magenta{($p$ = .397)}.
Moreover, interaction conditions had not effect on stimulation and attractiveness qualities \magenta{($p$ = .178 and $p$ = .639, respectively)}, which were rated with similar medium scores.
Regardless of the histograms, seeking peaks in energy variations provides only a basic user experience level.
Motion noise---which was noticed out loud by several participants---must also have degraded the user experience of physical shape-changing histograms.
However, results are more contrasted among the two other qualities with an effect of the interaction conditions \magenta{(all $p$ < .001)}.
CairnSCREEN is the only histogram reaching a high pragmatic quality \magenta{(all $p$ < .03)} but also the only histogram not reaching a high identification quality \magenta{(all $p$ < .001)}.

\begin{table}
  \caption{User Study -- User Experience (N=28).}
  \label{tab:results:study:ux}
  {\tablefontsize
    \begin{tabular}{
        @{\hspace{1pt}}
        p{50pt}
        @{\hspace{6pt}}p{38pt}@{\hspace{6pt}}p{38pt}@{\hspace{6pt}}p{38pt}@{\hspace{6pt}}p{38pt}@{\hspace{6pt}}p{38pt}
        @{\hspace{1pt}}
      }
      \midrule
     & \multicolumn{5}{l}{\tbh{AttrakDiff Scale}}\\
    \cmidrule{2-6}
    \textit{Histogram} & \textit{PQ}& \textit{HQI}& \textit{HQS}& \textit{ATT}& \textit{Overall}\tabularnewline
    \midrule
    PlantSCREEN  &\raL   {\tiny$^{\blacktriangle2}_{\circ}$}\hfill\mbox{-0.3}$\pm$0.3 &\raL   {\tiny$^{\blacktriangle1}_{\bullet}$}\hfill1.7$\pm$0.3 &\raL\bf{\tiny$^{\blacktriangle1}_{\bullet}$}\hfill1.4$\pm$0.3 &\raL\bf{\tiny$^{\blacktriangle2}_{\circ}$}\hfill1.2$\pm$0.2 &\raL   {\tiny$^{\blacktriangle2}_{\circ}$}\hfill1.0$\pm$0.2\tabularnewline
    PlantFORM    &\raL   {\tiny$^{\blacktriangle1}_{\circ}$}\hfill0.1$\pm$0.4 &\raL   {\tiny$^{\blacktriangle1}_{\bullet}$}\hfill1.8$\pm$0.4 &\raL   {\tiny$^{\blacktriangle1}_{\bullet}$}\hfill1.0$\pm$0.4 &\raL   {\tiny$^{\blacktriangle1}_{\circ}$}\hfill1.0$\pm$0.4 &\raL   {\tiny$^{\blacktriangle2}_{\circ}$}\hfill1.0$\pm$0.3\tabularnewline
    CairnSCREEN  &\raL\bf{\tiny$^{\blacktriangle1}_{\circ}$}\hfill1.9$\pm$0.2 &\raL   {\tiny$^{\blacktriangle1}_{\circ}$}\hfill\mbox{-0.2}$\pm$0.5 &\raL   {\tiny$^{\blacktriangle1}_{\circ}$}\hfill0.8$\pm$0.4 &\raL   {\tiny$^{\blacktriangle1}_{\circ}$}\hfill1.0$\pm$0.4 &\raL   {\tiny$^{\blacktriangle2}_{\circ}$}\hfill0.9$\pm$0.3\tabularnewline
    CairnFORM    &\raL   {\tiny$^{\blacktriangle2}_{\circ}$}\hfill1.0$\pm$0.4 &\raL\bf{\tiny$^{\blacktriangle1}_{\bullet}$}\hfill1.9$\pm$0.3 &\raL   {\tiny$^{\blacktriangle1}_{\circ}$}\hfill0.8$\pm$0.3 &\raL\bf{\tiny$^{\blacktriangle2}_{\circ}$}\hfill1.2$\pm$0.3 &\raL\bf{\tiny$^{\blacktriangle2}_{\circ}$}\hfill1.2$\pm$0.2\tabularnewline
  \midrule
   & \multicolumn{5}{l}{\textbf{\textit{p}-values}} \\
  \cmidrule{2-6}
  & \ts{(2)} & \ts{(2)} & \ts{(2)} & \ts{(2)} & \ts{(2)}  \\
  \midrule 
PS/PF/CS/CF & {\bf .000} ***   & {\bf .000} ***   &      .178        &      .639        &      .397        \\
    PS/PF       & {\bf .028} *     &      .304        & {\bf .030} *     &      .256        &      .476        \\
    PS/CS       & {\bf .000} ***   & {\bf .000} ***   & {\bf .033} *     &      .317        &      .328        \\
    PS/CF       & {\bf .000} ***   &      .136        & {\bf .026} *     &      .393        &      .101        \\
    PF/CS       & {\bf .000} ***   & {\bf .000} ***   &      .245        &      .389        &      .377        \\
    PF/CF       & {\bf .005} **    &      .313        &      .211        &      .259        &      .166        \\
    CS/CF       & {\bf .001} **    & {\bf .000} ***   &      .489        &      .264        & {\bf .014} *     \\
  \midrule
  \end{tabular}}
  {\parbox{\linewidth}{\scriptsize
      \textit{Notes.}
      \textbf{Format:} mean$\pm$margin.
      \textbf{AttrakDiff qualities:} `PQ': pragmatic quality, `HQI': hedonic quality - identification, `HQS': hedonic quality - stimulation, `ATT': attractiveness.
      \textbf{Scores} range from -3: ``Horribly bad,'' to 3: ``Extremely good.''
      \textbf{Values in bold print} are best scores and $p$-values under 0.05.
      \textbf{Outliers} out of 5th and 95th percentiles: $\vartriangle$~= No outliers; $\blacktriangle{n}$~= $n$ outliers. 
      \textbf{Distribution:} $\circ$~= Normal; $\bullet$~= Non-normal (by Shapiro--Wilk test).
      \textbf{Confidence intervals} are at a 95\% level from a normal distribution for normally distributed scores and from the percentile bootstrap method for non-normally distributed scores.
      \textbf{Acronyms:}  `PS' = PlantSCREEN, `PF' = PlantFORM, `CS' = CairnSCREEN, `CF' = CairnFORM.
      \textbf{Statistical tests:}
      (2) Friedman's ANOVA when four groups; Wilcoxon signed-rank test when two groups.
      \textbf{Statistical significance marks} are: \dg{}~$p<.10$, *~$p<.05$, **~$p<.01$, ***~$p<.001$.
    }
  }
\end{table}
 
\def\uxdetailWa{0.343} \def\uxdetailWb{0.209} \def\uxdetailSp{0pt}
\begin{figure*}
  \captionsetup[subfigure]{oneside,margin={2.1cm,0cm}}
  \begin{subfigure}[b]{\uxdetailWa\linewidth}
    \centering
    \includegraphics[width=1.0\linewidth]{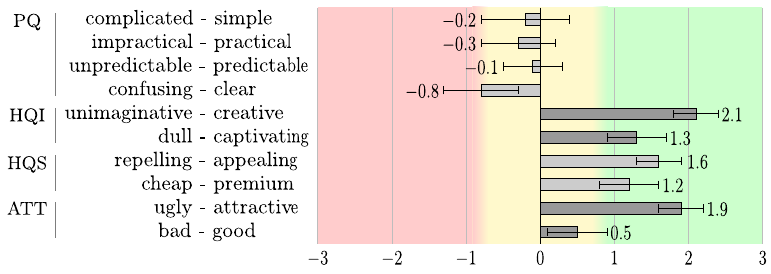}
    \caption{PlantSCREEN}
    \label{fig:results:study:ux_detail:plantscreen}
    \Description{The horizontal bar chart shows means and 95\% CIs of scores for the ten items of the AttrakDiff short questionnaire for PlantSCREEN.}
  \end{subfigure}
  \hspace{\uxdetailSp}
  \captionsetup[subfigure]{oneside,margin={0cm,0cm}}
  \begin{subfigure}[b]{\uxdetailWb\linewidth}
    \centering
    \includegraphics[width=1.0\linewidth]{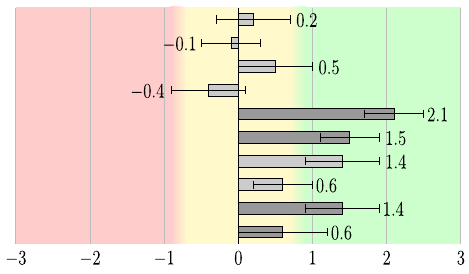}
    \caption{PlantFORM}
    \label{fig:results:study:ux_detail:plantform}
    \Description{The horizontal bar chart shows means and 95\% CIs of scores for the ten items of the AttrakDiff short questionnaire for PlantFORM.}
  \end{subfigure}
  \hspace{\uxdetailSp}
  \begin{subfigure}[b]{\uxdetailWb\linewidth}
    \centering
    \includegraphics[width=1.0\linewidth]{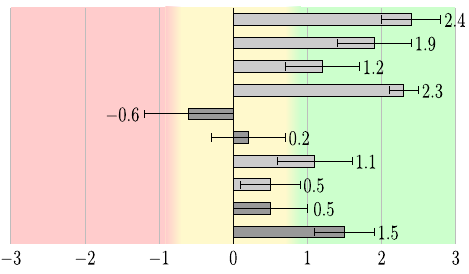}
    \caption{CairnSCREEN}
    \label{fig:results:study:ux_detail:cairnscreen}
    \Description{The horizontal bar chart shows means and 95\% CIs of scores for the ten items of the AttrakDiff short questionnaire for CairnSCREEN.}
  \end{subfigure}
  \hspace{\uxdetailSp}
  \begin{subfigure}[b]{\uxdetailWb\linewidth}
    \centering
    \includegraphics[width=1.0\linewidth]{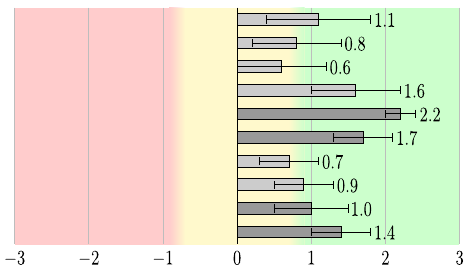}
    \caption{CairnFORM}
    \label{fig:results:study:ux_detail:cairnform}
    \Description{The horizontal bar chart shows means and 95\% CIs of scores for the ten items of the AttrakDiff short questionnaire for CairnFORM.}
  \end{subfigure}
    {\parbox{0.90\linewidth}{\scriptsize
      \textit{Notes.} \textbf{AttrakDiff qualities:} `PQ': pragmatic quality, `HQI': hedonic quality - identification, `HQS': hedonic quality - stimulation, `ATT': attractiveness. \textbf{Scores} range from -3: ``Horribly bad''; to 3: ``Extremely good.'' \textbf{Error bars} are 95\% confidence intervals from a normal distribution for normally distributed scores and from the percentile bootstrap method for non-normally distributed scores.
    }}
  \caption{User Study -- User experience through the ten items of the AttrakDiff short questionnaire (N=28).}
  \label{fig:results:study:ux_detail}
\end{figure*}

The semi-structured interview results on user preferences are reported in \autoref{tab:results:study:preferred}, and some user feedback on PlantFORM in \autoref{tab:userfeedback}.
The results reveal that PlantFORM is highly perceived as the closest to nature compared with the three other histograms.
Participants praised the use of wood and recyclable materials, the hand-work, and the perceived ``low-tech'' aspects.
They also appreciated that electronics are hidden and that no screens are used.
PlantFORM was also rated only the second most innovative and exciting prototype, for example, because they had never seen an interface of this kind before and found it ``intriguing.''

\begin{table*}
  \caption{User Study -- Participant rates over their global preferences (N=28).}
  \label{tab:results:study:preferred}
  {\tablefontsize
    \begin{tabular}{
        l
        @{}p{4pt}
        p{15pt}@{~}p{25pt}
        p{15pt}@{~}p{25pt}
        p{15pt}@{~}p{25pt}
        p{15pt}@{~}p{25pt}
        p{15pt}@{~}p{25pt}
        p{15pt}@{~}p{25pt}
        p{15pt}@{~}p{25pt}
      }
      \midrule
     && \multicolumn{14}{l}{\tbh{The most\ldots}}\\
    \cmidrule{3-16}
    \textit{Histogram} && \multicolumn{2}{l}{\textit{Nature-like}}& \multicolumn{2}{l}{\textit{Innovative}}& \multicolumn{2}{l}{\textit{Exciting}}& \multicolumn{2}{l}{\textit{Fascinating}}& \multicolumn{2}{l}{\textit{Motivating}}& \multicolumn{2}{l}{\textit{Adapted}}& \multicolumn{2}{l}{\textit{Overall}}\tabularnewline
    \midrule
    PlantSCREEN &&\raL     4\%&{\tiny [0\%,~16\%]}  &\raL    14\%&{\tiny [5\%,~30\%]}  &\raL    18\%&{\tiny [7\%,~35\%]}  &\raL    25\%&{\tiny [12\%,~43\%]} &\raL\bf 29\%&{\tiny\bf [14\%,~47\%]} &\raL     7\%&{\tiny [2\%,~21\%]}  &\raL    24\%&{\tiny [17\%,~33\%]}\tabularnewline
    PlantFORM   &&\raL\bf 93\%&{\tiny\bf [79\%,~98\%]} &\raL    25\%&{\tiny [12\%,~43\%]} &\raL    25\%&{\tiny [12\%,~43\%]} &\raL    29\%&{\tiny [14\%,~47\%]} &\raL    21\%&{\tiny [10\%,~39\%]} &\raL    14\%&{\tiny [5\%,~30\%]}  &\raL\bf 52\%&{\tiny\bf [43\%,~61\%]}\tabularnewline
    CairnSCREEN &&\raL     0\%&{\tiny [0\%,~8\%]}   &\raL     7\%&{\tiny [2\%,~21\%]}  &\raL    14\%&{\tiny [5\%,~30\%]}  &\raL     4\%&{\tiny [0\%,~16\%]}  &\raL\bf 29\%&{\tiny\bf [14\%,~47\%]} &\raL\bf 54\%&{\tiny\bf [36\%,~71\%]} &\raL    27\%&{\tiny [19\%,~36\%]}\tabularnewline
    CairnFORM   &&\raL     4\%&{\tiny [0\%,~16\%]}  &\raL\bf 54\%&{\tiny\bf [36\%,~71\%]} &\raL\bf 43\%&{\tiny\bf [26\%,~61\%]} &\raL\bf 43\%&{\tiny\bf [26\%,~61\%]} &\raL    21\%&{\tiny [10\%,~39\%]} &\raL    25\%&{\tiny [12\%,~43\%]} &\raL    47\%&{\tiny [38\%,~56\%]}\tabularnewline
  \midrule
  \end{tabular}}
  {\parbox{\linewidth}{\centering
      \scriptsize
      \textit{Notes.}
      \textbf{Format:} mean [lower bound, upper bound].
      \textbf{Values in bold print} are the best rates.\\
      \textbf{Confidence intervals} are at a 95\% level, from Jeffreys Bayesian method.
    }
  }
\end{table*}

\def\colwidth{14.8cm}
\begin{table}
  \centering
  \caption{Some user feedback on PlantFORM.}~\label{tab:userfeedback} {\tablefontsize
    \begin{tabular}{p{10pt}p{\colwidth}} \midrule
    \multicolumn{2}{l}{\textbf{\textit{Performance of Energy Shift Tasks}}} \\
    \midrule
     {P4:} & \textit{``[\ldots] the design of the end of the leaves does not clearly identify peaks in availability rates.''} \\
     {P1:} & \textit{``[\ldots] lack of light information to differentiate ascending and descending phase.''} \\
{P6:} & \textit{``Easy to use.''} \\
    \midrule
    \multicolumn{2}{l}{\textbf{\textit{User Experience and Preferences}}} \\ 
    \midrule
     {P9:} & \textit{``Mechanical noise to be reduced.''} \\
     {P8:} & \textit{``It is good that it is noisy because we know when it refreshes. It can be a defect as well as an advantage.''} \\
     {P21:} & \textit{``Leaves that can grow instead of just unfurling.''} \\
{P20:} & \textit{``PlantFORM would win the prize for sobriety, although it may not be the case with the electronics behind it, but [PlantFORM] may not work continuously and be continuously powered.''} \\
     {P7:} & \textit{``PlantFORM [is the most motivating one] because of its conception, hidden cables, and materials choices. [\ldots] We are talking about renewable energy; it affects me enormously, so all this Plexiglass [of CairnFORM] [\ldots], if we are in renewable energy, we must go all the way in the product we offer.''}\\ 
    \midrule
  \end{tabular}
  }
\end{table}

CairnFORM was rated the most innovative and exciting prototype because of the use of LEDs, ring motion, and translucent PMMA that make technology visible.
Participants also found this prototype ``impressive,'' ``playful,'' and ``geek,'' and one compared it to ``a building.''
The two most fascinating prototypes were CairnFORM and PlantFORM because of the physical motion of the rings and leaves.
For example, one participant said: \textit{``[PlantFORM] recalls the natural movement of a plant''} (P9).

Assumption A9 on materials' effect on tech-trend perception holds.
Indeed, PlantFORM is perceived as the most nature-like by most participants, whereas PlantSCREEN by only a few.
Therefore, even if these two prototypes implement the same plant-like design and even if PlantSCREEN uses a wooden case, conveying data through graphical monitors increases the distance to nature compared with physical cardboard and thick paper.
Moreover, bar-like charts, which convey data through graphical monitors or physical plastic, are perceived as far from nature.

\section{Discussion}\label{sec:discussion}

This section highlights the contributions of this research to Human--Plant Interaction, Eco-Forecasts, Embellished Charts, and Shape-Changing Interfaces.

\subsection{Benefits to Human--Plant Interaction and Eco-Forecasts}

This research extends the literature of Human--Plant Interaction and Eco-Forecasts by implementing and evaluating a new physical artifact and a new shape-changing mechanism usable for plant-like prototyping.

Metaphors of plants are used to represent or convey various kinds of data (e.g., counters \cite{chien2015biogotchi,degraen2019overgrown}, gauges \cite{antifakos2003laughinglily,holstius2004infotropism,hong2015better,seow_pudica_2022}, and bitmaps \cite{gentile2018plantxel,takaki2014mossxels,tanaka2024programmablegrass}). However, data series are still to be explored in depth, for example, to represent environment-related data that evolves gradually (e.g., over hours \cite{daniel2019cairnform,schrammel_forewatch_2011,kjeldskov_eco_forecasting_2015}, days \cite{costanza_doing_laundry_2014,bourgeois_conversations_2014}, or both \cite{simm_tiree_2015}). Indeed, nature already inspired several environmental visualizations \cite{chalal_visualisation_2022}. However, our results show that maximizing naturalness and readability compels compromise that restricts the design possibilities of plant-like charts. Whereas curvy and straight trunks reach similar readability rates, alternated anchorings bring higher aesthetics than one-sided anchorings but lead to slightly lower readability and clarity. Finally, our design space exploration leads to a single possible design to encode data series through plants: curvy two-sided vertical charts with unfurling leaves.

We implemented this design solution through a physical prototype by introducing a new mechanism for unfurling leaves, which is made of an extension spring, a pulling cable, and a guiding structure.
The users felt the bar-like prototypes were more adapted than plant-like prototypes to answer the repeated questions, even if the plant-like charts were rated more appealing and attractive. Moreover, high aesthetics should contribute to successfully integrating ambient charts into users' daily environment. Finally, the physical plant-like prototype was rated as the closest to nature. This prototype should best act as a vector of normative influence \cite{oinas-kukkonen_persuasive_2009} to constantly recall the environmental purpose of eco-forecasting.

\subsection{Benefits to Embellished Chart}

A previous study on bar shapes showed decreased accuracy when comparing rounded endings with traditional square bars (measured by the difference between judged value and true value) on charts of three values, especially on a relative comparison task of two bars \cite{skau2015evaluation}, which relates to our task of reading variations' patterns (i.e., retrieving variations' start, peak, and end, on charts of ten values). Our results also found decreased success rates on the rounded shapes of leaf and pebble charts, compared with bar charts with clear endings---however, our task results in smaller gaps.

Furthermore, success rates also decreased when tiny decorations were added along bars. Successful bar comparisons must involve reading bars' lengths clearly in addition to the relative positions of bar endings. However, only decorations on bars' outsides may decrease length reading because a previous study on inside pictorials had no discernible impact \cite{skau2017readability}..

User preferences rated the bar-like chart prototypes as simpler, clearer, and more practical than the leaf chart prototypes, especially when displayed graphically. The users also felt the bar-like charts were more adapted to peak hour and slope endings retrieval tasks. However, the plant-like chart prototypes were rated more appealing and attractive, which confirms the better aesthetics of the leaf chart design. These preferences align with previous results of users who feel plain charts are clearer and faster to read \cite{andry_interpreting_2021,bateman2010junk} but like embellished charts because they are more aesthetically pleasing \cite{andry_interpreting_2021}.

In addition, our work studies encoding through folded shapes, alternatively to the more traditional encoding through shapes that are extended \cite{andry_interpreting_2021,skau2015evaluation,skau2017readability} or repeated \cite{burns2022pictographs,haroz2015visualization}. Such data encoding is quite unusual, so we assessed the learnability of rate encoding through leaf unfurling. Even if the presence of labels---at leaf anchorings along the trunk that stands as an axis---recall a chart \cite{arunkumar2024image}, the scope of the study and tasks were prior indications informing users that the prototypes were encoding data series. However, no explanations were given about rate encoding through unfolding motion. The results show that a short learning phase is required and that the horizontal line is implicitly understood as the maximum value. Thereby, ambient charts using folded shapes that are installed by a third party in public or semi-public spaces ought to be understood without explanations nor a scale of values, as long data are rates ranging from zero to a maximum value.

\subsection{Benefits to Shape-Changing Interfaces}

The choice of homogeneous materials emphasizes the perceived tech trends. Indeed, whereas wood, cardboard, and thick paper materials maximize the low-tech aspects and the perceived naturalness of a physical plant-like prototype, the high-tech aspects of the physical prototype stacking illuminated PMMA-made rings were perceived as the most innovative. These results align with some marketing studies on the perception of packaging materials \cite{lindh2016consumer,nguyen2020consumer}. Moreover, prototypes with half-combined aspects, melting wooden cases and LCD screens received only low innovation and naturalness perceptions.

The user study results confirm, with bar-like charts, previous work that found similar readability between physical and graphical bar charts \cite{jansen_evaluating_2013}. However, a superiority of the physical chart was found for plant-like charts. The implementation flaws of PlantSCREEN must explain the difference in readability (i.e., excessive unfurling). Therefore, we believe that plant-like charts should also reach similar reading qualities even if not confirmed by our physical and graphical prototypes.

\section{Limitations}\label{sec:limitations}

The prototypes of the user study prevented the direct drawing of definitive conclusions (e.g., on success rates and preference for the task), because of some implementation flaws or non-fully equivalent encoding.
For instance, a zero rate was encoded by CairnSCREEN with an empty position, whereas shapes of the three other prototypes were always present.
Moreover, unfurling leaves beyond the horizontal line downgraded the performance of PlantSCREEN.
The same, the furling of the zero position of PlantFORM was less emphasized than on PlantSCREEN, and granularity between the eleven leaf positions was also less discriminative than six.

Furthermore, folding principles are still to be explored deeper and improved (e.g., shape memory alloy \cite{cheng2014mood-fern}) to enable both full shape folding for zero rates encoding (i.e., PlantHISTO in \autoref{fig:online2:histograms:planthisto}) and motion discretization for efficient granularity of data encoding (i.e., the six positions in \autoref{fig:design_inspiration:ornementation}).

\section{Conclusion}\label{sec:conclusion}

This paper frames a design space of plant-like charts, leading to one possible design maximizing readability and aesthetics: a curvy trunk with a two-sided (axis-symmetric) anchoring of unfurling leaves. The physical implementation of such a chart through shape change was done by introducing a new folding mechanism. However, the readability of our leaf chart prototypes remained lower than for the bar-like prototypes. Overcoming some implementation flaws should attenuate this difference and make plant-like charts usable for energy management, such as consumption shift, that only requires peak and slope retrievals. Indeed, natural fibers (such as wood, cardboard, and thick paper) increased the perception of naturalness, reinforcing the purpose of acting for the environment. Moreover, the higher aesthetics of the plant-like design should better integrate into users' environment to display energy forecasts daily and ambiently.

This research confirms the previous results on the higher aesthetics but lower clarity and decreased accuracy of embellished charts compared with bar and bar-like charts. It also deepens our understanding of the reading of bar shapes because of decreased readability when tiny decorations along bars prevent reading their length well and comparing their endings' positions. It also shows that a short learning phase is required when encoding rates through unusual folded shapes and that the horizontal line is implicitly understood as the maximum value. Therefore, encoding data series through plant-like charts is possible as long data are rates that range from zero to a maximum value.

\ifFINAL
\begin{acks}
  The authors thank all the researchers who were involved in the previous steps of the project, which this work builds upon:
  Stephane Kreckelbergh, who provided knowledge on micro-grids and renewable-energy contexts, which stands as the basement of this work,
  as well as Nadine Couture and Maxime Daniel, who participated in several ideation sessions when designing the energy practice.
  The authors also thank all the internship students involved during the design and implementation of PlantFORM and PlantSCREEN:
  Liangwei Zhao for preparing the electronic components and taking part in the brainstorming, design, and fabrication of the first version of PlantSCREEN and PlantFORM wooden casings;
  Brieuc Mandin for preparing the 3D files and assembling the parts of the second version of PlantSCREEN, and for improving the wiring of PlantFORM's electronics;
  Baptiste Braun for preparing and doing the wiring of the electronic boards of PlantFORM;
  Mathieu Romain for co-designing the prototype, preparing the 3D files and assembling the part of PlantFORM.
  The authors also especially thank Maxime Daniel for allowing the reuse of CairnFORM and CairnSCREEN hardware for the needs of the comparative user study. Subsequently, the authors also thank Adrien Fat Cheung for the design of labels fixing system of CairnFORM and Gauthier Capdepon de Bigu for the redesign of CairnFORM's rings' internal wiring.  
  Last but not least, the authors thank all the reviewers from ACM TEI 2024 and ACM IMWUT who helped restructure the outline and the message of this paper.
\end{acks}
 \fi

\bibliographystyle{ACM-Reference-Format}
\bibliography{paper.bib} 

\end{document}